\documentclass[12pt]{article}
\usepackage{amsmath,graphicx}
\usepackage[latin1]{inputenc}
\usepackage{hyperref}
\usepackage{xcolor}
\usepackage{booktabs}
\usepackage{longtable}
\DeclareMathOperator\Li{Li}

\def\l({\left(}
\def\r){\right)}

\begin{document}

\title{{\bf Fourier series for the three-dimensional random flight}}
\author{
Ricardo Garc\'\i a-Pelayo\thanks{%
E-mail: r.garcia-pelayo@upm.es} \\
\\
ETS de Ingenier\'ia Aeron\'autica \\
Plaza del Cardenal Cisneros, 3 \\
Universidad Polit\'ecnica de Madrid \\
Madrid 28040, Spain\\}
\date{}
\maketitle

\maketitle

\begin{abstract}
The probability density function of the random flight with isotropic initial conditions is obtained by an expansion in the number of collisions and the in the spatial harmonics of the solution, as in a Fourier series. The method holds for any dimension and is worked out in detail for the three dimensional case. In this case the probability density functions conditional to 1 and 2 collisions are also found using a different method, which yields them in terms of elementary functions and the polylogarithm function Li$_2$. The latter method is exact in the sense that one does not have to truncate a series, as in the first method. This provides a reference to decide where to truncate the series. A link is provided to a web page where the reader may download the series truncated at 132 collisions; for times larger than 100 times the average inter-collision time, the Gaussian approximations is used. The case in which the initial condition is a particle moving along a fixed direction is briefly considered.
\end{abstract}

\centerline{}

\tableofcontents

\section{Introduction and applications}

In the random flight a particle moves with a constant speed $v$ until, at times uniformly distributed with density $\lambda$, it takes a uniformly distributed direction, while maintaining the same speed. To summarize it colorfully, ``a photon in a box of mirrors'' \cite{Serva2021}. There are, of course, generalizations of this definition (see page xix of \cite{Kolesnik2021}). The random flight is the simplest case of ``continuous time random walk" (CTRW) (\cite{Montroll1965}, p. 177). The words ``random flight" may also be used when the changes of direction do not take place at uniformly distributed times but at equally spaced times. This usage is due to Rayleigh \cite{Rayleigh1919} (reprinted in \cite{Rayleigh1964}) as quoted in page 55 of \cite{Hughes1996}. See references \cite{Hughes1996} (chapters 2 and 5) and \cite{RGPN2012cap,Weiss2002,Kolesnik2021,RGPN2023} and references therein for a broader introduction to the subject.

The random flight with isotropic initial conditions was solved in 1 dimension by Goldstein in 1950 \cite{Goldstein1951}, in 2 dimensions in 1987 by Stadje \cite{Stadje1987} and in 1993 by Masoliver et alii \cite{Masoliver1993}, in 4 dimensions by Orsingher and De Gregorio in 2007 \cite{Orsingher2007} and in 6 dimensions by Kolesnik in 2009 \cite{Kolesnik2009}.

There have been attempts to formulate the random flight in more than 1 dimension as a ``persistent random walk'', in the sense of a multidimensional telegraph equation, but they have failed \cite{Masoliver1993, Kolesnik2011}.

The three-dimensional random flight is a Lorentz gas \cite{Lorentz1905} in the limiting case of point-like scatterers. The three dimensional case has been solved in the Fourier-Laplace space \cite{McKean1967,Case1967,Hauge1970}. As a Lorentz gas the topic of this article is still being researched \cite{Zeitz2017,Biroli2021}. More recently the problem has been posed in a biological setting, where it is known as the run-and-tumble model for the diffusion of bacteriae \cite{Martens2012, Gradenigo2019, Proesmans2020, Mori2021a, Mori2021b}. Another applications is seismic waves \cite{Hoshiba1991,Sato2012}.

An application of the three-dimensional random flight solved in this article which comes quickly to mind (see the second sentence of this article) is the scattering of photons \cite{Durian1997,Lemieux1998,Burshtein1988,Kofman1990,Miri2005a,Joensson2020}. An interesting particular case of scattering of photons is the zodiacal light \cite{Leinert1975,Hahn2002}. The dust which causes zodiacal light has a size of 10-300 microns \cite{Peucker-Ehrenbrink2001}, that is, well above the wavelength of visible light (0.38 to 0.75 microns). Therefore the reflection of light follows the laws of geometrical optics. This, on a sphere, leads to isotropic scattering (see, e. g., footnote of p. 263 of reference \cite{Tousey1957}) and thus the results of this article are pertinent.

The random walk solved here is also the random walk which supposedly accounts for the Lambert reflection \cite{Zook1976}. This kind of random walk is used for visual rendering of objects \cite{Earp2007,Wann2023}.

Like any analytic result on random processes, the results presented here can be used to test the accuracy of simulation programs. See reference \cite{Martelli2021} for a recent discussion of this topic.


The random flight problem presented here can be cast as a Boltzmann equation for the one-particle distribution, as in equation (4) of reference \cite{Hauge1970}, with its $\pi \rho v a^2$ substituted by our $\lambda$. There are exact solutions to this Boltzmann equation in the Fourier-Laplace space-time \cite{McKean1967,Hauge1970,Cornu2006}, but no exact solutions in the space and time variables. The solution given here, which admits arbitrary initial conditions as explained in the Conclusions, should be a solution to its corresponding one-particle distribution Boltzmann equation. Exploring the relation between both displays of the same physical problem might furnish a deeper understanding of the corresponding Boltzmann equation.

Some remarks which are better understood after having read the article are presented in the Conclusions section.

\section{Organization and results}

Some preliminary definitions, and an expansion for the random flight's pdf and its properties are presented in sections \ref{Definitions} and \ref{Collision expansion and scaling relations}, respectively. The main result of this article is that the probability density function (henceforth, pdf) of the random flight with isotropic initial conditions can be obtained by an expansion in the number of collisions and the spatial frequencies of the solution, as in a Fourier series. This is explained in sections \ref{Solutions as Fourier series} and \ref{Fourier series of r and I}. This method is applicable to any dimension, including non-integer dimensions.

While the methods of this article can be applied to any number of dimensions, it is the three dimensional case - conspicuously missing from the list of solutions given in the second paragraph of this article - the one which is developed here. The result, which we display in this paragraph, is the pdf $\rho_S$ of the position of a 3-dimensional random flight with isotropic initial conditions, with speed $v$ and with $\lambda$ collisions per unit time. This pdf is the sum of an expanding spherical shell of norm $e^{-\lambda t}$ and a pdf $(1 - e^{-\lambda t}) \rho(\vec r, t)$ of norm $1 - e^{-\lambda t}$, that is,

\begin{equation}\label{decomposition.0}
\rho_S(\vec r, t) = e^{-\lambda t} {\delta(r-vt) \over 4 \pi r^2} + (1 - e^{-\lambda t}) \rho(\vec r, t),
\end{equation}
with $\rho$ normalized to 1. The collision expansion (used in reference \cite{RGPN2023} and called Born expansion in references \cite{RGPN1995,RGPN2012cap}) for the pdf $\rho$ is

\begin{equation}\label{PoissonExpansion.0}
\rho(\vec r, t) = {1 \over e^{\lambda t} - 1} \sum_{c=1}^\infty {(\lambda t)^c \over c!} \rho_{c}(\vec r, t),
\end{equation}
where $\rho_{c}$ is the pdf conditional to $c$ collisions. The two previous formulae are easy to understand. In sections \ref{Solutions as Fourier series} and \ref{Fourier series of r and I} of this article the solution by Fourier series used in section 3 of reference \cite{RGPN2023} to solve the one-dimensional case is extended to the three-dimensional case. We obtain the following analytic form for $\rho_{c}\ (c=1,2,3,...):$

$$
\rho_{c}(\vec r,t) =
$$

\begin{equation}\label{ro1sc.1.0}
{1 \over v t}\ \l( {\langle r^{-2} \rangle_{\rho_c}(1/v) \over 4 \pi (vt)^2} + {1 \over 2 \pi} \sum_{h=1}^{\infty} \left[ {\langle r^{-2} \rangle_{\rho_c}(1/v) \over (vt)^2} + {c! \over (vt)^2} \sum_{m=1}^\infty { (-\pi^2)^m\ C(c,m-1) \over 2 m (2m-2+c)!} \cdot h^{2m} \right] \cos {\pi h \over v t} |\vec r| \r),
\end{equation}
where
\begin{equation}\label{C(c,m)def.0}
C(c,m) \equiv \sum_{\underset{i_1 + \cdots + i_{c+1} = m} {i_1, \cdots, i_{c+1} \in N}} {1 \over (2 i_1+1) \cdots (2 i_{c+1}+1)}
\end{equation}
and

\begin{equation}\label{limFS.0}
\langle r^{-2} \rangle_{\rho_c}(1/v) = - \lim_{h \to \infty} c! \sum_{m=1}^\infty { (-\pi^2)^m\ C(c,m-1) \over 2 m (2m-2+c)!} h^{2m}.
\end{equation}

In Appendix C the computation of these expressions and their errors will be discussed and it will be shown that the Fourier series \eqref{ro1sc.1.0} works, from a numerical point of view, quite well for $c=3,4,5,...$. But in the three dimensional case the main method presented in this article is numerically not that good when used to find the probability density function conditional to 1 and 2 collisions. To find these two pdf's a different method is applied (in section \ref{Exact solution} for the 1 collision case and in Appendix B for 2 collisions) which is obtained by developing a method presented in reference \cite{RGPN2012}. These two pdf's also provide a standard to test the main method, because there are closed analytical expressions for them and therefore they are ``exact'' with respect to expressions involving sums which have to be truncated. The expression for $\rho_1$ is quite simple:

\begin{equation}\label{rho1.0}
\rho_1(\vec r,t) = {1 \over 4 \pi (vt)^2 |\vec r|} \ln {v t+|\vec r| \over v t-|\vec r|}.
\end{equation}
The analytic expression for $\rho_2$ is very long and is obtained in Appendix B.

In section \ref{Graphs} some graphs are displayed which show how the probability density of the three-dimensional pdf expands with time. The match with simulations is also addressed there.

The reader may download software to compute the pdf $\rho(\vec r,t)$ \cite{RGPNgooglesite}. This software is written in a high-level programming language, so that readers who don't have that particular software may at least read it. See the beginning of section \ref{Graphs} for further details.

For the reader who has read reference \cite{RGPN2023}, we note that there is a subtle difference between the random flight in one dimension and the random flight in more than one dimension. In one dimension if a particle is scattered there is a probability of 1/2 that it keeps its direction, while in the latter case this probability is 0. Therefore in more than 1 dimension the number of scatterings is equal to the number of changes of directions with probability 1, and the distinction made between the two models shown in section 2 of \cite{RGPN2023} disappears.

\section{Definitions}\label{Definitions}

The pdf $\rho_S$ of the position of a 3-dimensional random flight with isotropic initial conditions, with $\lambda$ collisions per unit time and speed $v$ is the sum of an expanding spherical shell of norm $e^{-\lambda t}$ and a pdf of value $(1 - e^{-\lambda t}) \rho(\vec r, t)$ and norm $1 - e^{-\lambda t}$, that is,

\begin{equation}\label{decomposition}
\rho_S(\vec r, t) = e^{-\lambda t} {\delta(r-vt) \over 4 \pi r^2} + (1 - e^{-\lambda t}) \rho(\vec r, t),
\end{equation}
with $\rho$ normalized to 1. Henceforth, except for two mentions of $\rho_S$ in subsection \ref{ssec:Moments} and in the Conclusions, we deal exclusively with $\rho$ and its relatives, which we introduce right now:

\begin{equation}\label{eq:definitions}
\left.
\begin{array}{c}
\rho(\_,t): B_{(\vec 0, vt)} \longrightarrow \Re^{+} \\
\rho_r(\_,t): [0,vt] \longrightarrow \Re^{+} \\
\rho_I(\_,t): [0,vt] \longrightarrow \Re^{+} \\
\rho_{rs}(\_,t): [-vt,+vt] \longrightarrow \Re^{+} \\
\rho_{Is}(\_,t): [-vt,+vt] \longrightarrow \Re^{+} \\
\rho_{proj}(\_,t): [-vt,+vt] \longrightarrow \Re^{+} \\
\end{array}
\right\},
\end{equation}
where $B_{(\vec 0, vt)}$ is the ball centered at the origin and of radius $vt$. The function $\rho(\_,t)$ is our primary goal, although the radial distribution, $\rho_r(\_,t)$, may also be of interest. They are related as follows:

\begin{equation}\label{definition rho_r}
\rho_r(r,t) \equiv 4 \pi r^2 \rho(\vec r,t),
\end{equation}
where $\vec r$ is any point such that $|\vec r| = r$. The pdf $\rho_I(\_,t)$ is simply $\rho(\_,t)$, but regarded as a function of the modulus of $\vec r$:

\begin{equation}\label{rhoI}
\rho_I(r,t) \equiv \rho(\vec r,t),
\end{equation}
where $\vec r$ is any point such that $|\vec r| = r$. The pdf's $\rho_{rs}(\_,t)$ and $\rho_{Is}(\_,t)$ are very similar to $\rho_r(\_,t)$ and $\rho_{I}(\_,t)$, respectively, but they are more convenient for technical reasons, because they are symmetric about the origin and thus their Fourier series are just cosine series. Their definitions are

\begin{equation}\label{rhors}
\rho_{rs}(x,t) \equiv {1 \over 2} \rho_r(|x|,t) =
\left\{
\begin{array}{cc}
  {1 \over 2} 4 \pi x^2 \rho_I(x,t), & x \geq 0 \\
    &   \\
  {1 \over 2} 4 \pi x^2 \rho_I(-x,t), & x < 0
\end{array}
\right\}
\end{equation}
and

\begin{equation}\label{rhoIs}
\rho_{Is}(x,t) \equiv {1 \over 2} \rho_I(|x|,t).
\end{equation}
%

The pdf $\rho_{proj}(\_,t)$ is the projection of $\rho(\_,t)$ onto a diameter. Its connection with $\rho(\_,t)$ is more complicated than the above relations and it will be presented in section \ref{Exact solution}.
\bigskip

\section{Collision expansion and scaling relations}\label{Collision expansion and scaling relations}

First, a word on notation. For the convolution of two real functions of the real variable $x$, say functions $f$ and $g$, we use the notation $f \otimes g$. But when these depend also on other arguments, say $x_f$ and $x_g$, different from $x$, then we use a different notation, which we explain by the following example:

\begin{equation}\label{}
\big( f(x_f,\_) \otimes g(x_g,\_) \big)(x) \equiv \int_{-\infty}^{+\infty} dx'\ f(x_f,x') \otimes g(x_g,x-x').
\end{equation}

The collision expansion (used in reference \cite{RGPN2023} and called Born expansion in references \cite{RGPN1995,RGPN2012cap}) for the pdf $\rho$ is

\begin{equation}\label{PoissonExpansion}
\rho(\vec r, t) = {1 \over e^{\lambda t} - 1} \sum_{c=1}^\infty {(\lambda t)^c \over c!} \rho_{c}(\vec r, t),
\end{equation}
where $\rho_{c}$ is the pdf conditional to $c$ collisions. These conditional probabilities have been a topic of research in their own right \cite{MFranceschetti2007}. The expansion follows from the fact that the number of collisions in the time interval $[0,vt]$ is Poisson distributed. Note that the sum starts at $c=1$ because, as explained in eq. {\eqref{decomposition}, we are not including the expanding spherical shell, which is

\begin{equation}\label{}
\rho_{0}(r,t) = {\delta(r-vt) \over 4 \pi r^2}.
\end{equation}
It follows from expansion (15) in reference \cite{RGPN1995} or from expression (24) in \cite{RGPN2012cap} that

\begin{equation}\label{roc}
\rho_{c}(\vec r, t) = {c! \over t^c} \rho_{0}^{\otimes c+1}(\vec r, t).
\end{equation}
The exponential notation with the convolution sign $\otimes$ is akin to the usual exponential notation, that is $\rho_{0}^{\otimes 0}(x,t) = \delta(x,t)$, $\rho_{0}^{\otimes 1}(x,t) = \rho(x,t)$, $\rho_{0}^{\otimes 2}(x,t) = (\rho_{0} \otimes \rho_{0})(x,t)$, $\rho_{0}^{\otimes3}(x,t) = (\rho_{0} \otimes \rho_{0} \otimes \rho_{0})(x,t)$,... The convolution $\otimes$ is here a convolution on all the variables of $\rho_{0}$, that is, a convolution on space and time. For example,

\begin{equation}\label{}
\rho_{0}^{\otimes 3}(\vec r, t) = \int_0^{t_{2}} dt_1 \int_{0}^{t} dt_{2} \big( \rho_{0}(\_, t_1) \otimes_s \rho_{0}(\_, t_2-t_{1}) \otimes_s \rho_{0}(\_, t-t_2) \big) (\vec r),
\end{equation}
where the inner $\otimes_s$'s are convolutions on the space coordinates only, in consistence with the notation explained at the beginning of the section. The prefactor ${c! \over t^c}$ in equality \eqref{roc} ensures the normalization of $\rho_{c}$.

We denote by $\rho_{rs,c}$ the one-dimensional pdf $\rho_{rs}$ conditional to $c$ collisions. For this pdf the collision expansion reads

\begin{equation}\label{_{rs}.3}
\rho_{rs}(x,t) = {1 \over e^{\lambda t} - 1} \sum_{c=1}^\infty {(\lambda t)^c \over c!} \rho_{rs,c}(x,t).
\end{equation}
Similar expansions hold for the other pdf's defined at the beginning of this section.
\bigskip

An important feature of the pdf's conditional to the number of collisions, such as $\rho_{c}$ and $\rho_{rs,c}$, is that their dependence on time is just a scaling. It follows from the first of definitions \eqref{eq:definitions} and from expansion \eqref{PoissonExpansion} that the pdf $\rho_{c}(\_,v t)$ is a function which maps the sphere of radius $v t$ into $\Re^+$. Since the $\rho_c$'s are normalized, the quantity $\rho_{c}(\vec r,t) d^3r$ is the fraction of broken lines of length $v t$, with $c$ uniformly distributed corners and with uniformly distributed angles at the corners, which, starting at the origin of coordinates, end in a neighborhood $d^3r$ around $\vec r$. A similar definition holds for the quantity $\rho_{c}(\vec r',t') d^3r'$, where space and time have been scaled by the same ratio (see Fig. \ref{Scaling}).
There is a trivial one-to-one correspondence between the two sets of broken lines, related by a similarity transformation of ratio $t'/t$. Clearly (see Fig. \ref{Scaling}),

\begin{equation}\label{}
\rho_{c}(\vec r,t) d^3r = \rho_{c}(\vec r',t') d^3r' = \rho_{c}\Big( {t' \over t} \vec r, t' \Big) \Big({t' \over t}\Big)^3 d^3r,
\end{equation}
which implies the following scaling relation:

\begin{figure}[h]
\centering
  \includegraphics[width=0.75 \textwidth]{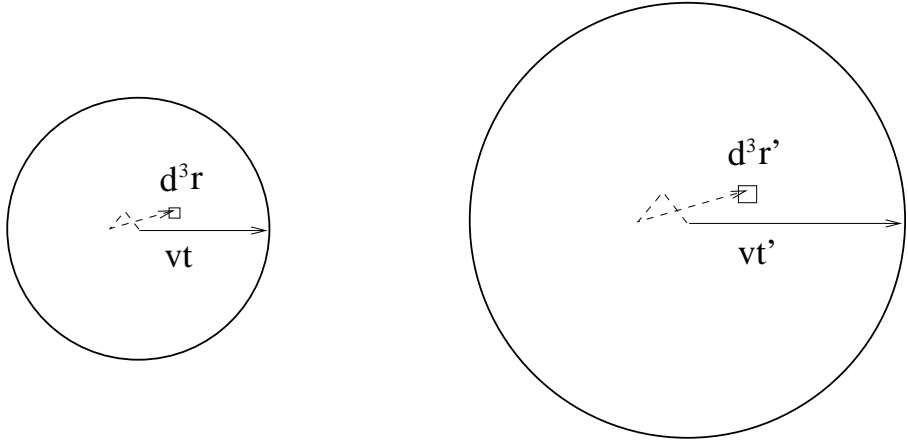}
  \caption{To the left, a broken line of length $vt$ starting at the origin and finishing at the element of volume $d^3r$ is shown. To the right, the corresponding broken line of length $vt'$ starting at the origin and finishing at the element of volume $d^3r'$ is shown. In both cases $c=2$.}\label{Scaling}
\end{figure}

\begin{equation}\label{ScalingRhoc1}
\rho_{c}(\vec r, t) = \Big( {t' \over t} \Big)^3 \rho_{c}\Big( {t' \over t} \vec r, t' \Big)
\ \overset{vt'=1}{\Longrightarrow} \
\rho_{c}(\vec r, t) = {1 \over (vt)^3} \rho_{c} \Big({\vec r \over vt}, {1 \over v} \Big).
\end{equation}
It follows from definition \eqref{definition rho_r} that $\rho_{r,c}$ scales as

\begin{equation}\label{ScalingRhoc2}
\rho_{r,c}(x, t) = {t' \over t}\ \rho_{r,c}\Big( {t \over t'} x, t' \Big)
\ \overset{vt'=1}{\Longrightarrow} \
\rho_{r,c}(x, t) = {1 \over vt} \rho_{r,c} \Big({x \over vt}, {1 \over v} \Big).
\end{equation}
This scaling relation also holds for $\rho_{r,c}$, $\rho_{rs,c}$, $\rho_{I,c}$ and $\rho_{Is,c}$.
%

The scalings are very good news because they show that the collision expansions give functions of two variables, $(x,t)$ or $(\vec r,t)$, as a linear combination of effectively one dimensional functions. This is so because the time dependence is just a scaling and the spatial dependence reduces to a dependence on just one variable: the radius.

As we shall see,
scaling relations \eqref{ScalingRhoc1} and \eqref{ScalingRhoc2} are used to speed the computation.
\bigskip

\section{Exact solution}\label{Exact solution}

In 1905 Pearson \cite{Pearson1905} posed the random flight problem but with changes of direction taking place not at random times but at regular intervals. In reference \cite{RGPN2012} we found the solution to the Pearson problem for odd dimensions. We explain now the method by which it was found because this method may be generalized to the current case of changes of direction taking place at Poisson distributed times. This generalization furnishes an analytic solution for up to two collisions. For three or more collisions the method becomes analytically untractable.

Since the convolution and the projection commute (section 9 of \cite{Kingman1963}, section 10 of \cite{RGPN2012}), in reference \cite{RGPN2012} the problem was projected onto one dimension, it was solved there, and the solution was projected back onto three dimensions. ``Projecting back from one onto three dimensions'' is something which has probably struck the reader as strange or plain wrong. Bear in mind, though, that the distributions involved in this problem are isotropic, and thus the inversion of the projection is possible. It can be achieved by a generalization of the Abel inverse transform. The projection and the generalized inverse Abel transform are given by formulae (17) and (18), respectively, of reference \cite{RGPN2012}. These formulae, written for the problem at hand, are:

\begin{equation}\label{Abel}
\rho_{proj}(x,t) = 2 \pi \int_x^\infty dx'\ \rho_I(x',t) x'
\end{equation}
which is the projection onto the diameter (not used in the current article, it is given here just for completeness)
and
\begin{equation}\label{inverseAbel.1}
\left.
\begin{array}{ccc}
\rho_{I}(x,t) & = & -{1 \over 2 \pi x} {\partial \over \partial x} \rho_{proj}(x,t) \\
 \\
\rho_{rs}(x,t) & = & -x {\partial \over \partial x} \rho_{proj}(x,t) \\
\end{array}
\right\},
\end{equation}
which is the generalized inverse Abel transforms for the case at hand.

We now make the preceding paragraph more concrete so that, hopefully, the reader might understand this article without reading reference \cite{RGPN2012}. The pdf $\rho_c(\_,t)$ after $c$ collisions which take place at equally spaced times $t \over c+1$,...,$c t \over c+1$ is obtained by convoluting $c+1$ expanding spherical waves, that is, $\rho_c(\_,t) = \rho_0 \big( \_,{t/(c+1)} \big)^{\otimes_s c+1}$, where $\rho_0(r,t) = {\delta(r-v t)/(4 \pi r^2)}$ (the subindex 0 stands for 0 collisions). We define the projection operator $P: \Re^3 \rightarrow \Re$ on isotropic functions and its inverse $P^{-1}$ by equations \eqref{Abel} and \eqref{inverseAbel.1}, respectively. Then using the fact that the convolution and the projection commute,

\begin{equation}\label{solution to Pearson}
\rho_c(\_,t) = P^{-1} P \l( \rho_0 \bigg( \_,{t \over c+1} \bigg)^{\otimes_s c+1} \r) =
P^{-1}\l( \rho_{proj,0} \bigg( \_,{t \over c+1} \bigg)^{\otimes_s c+1} \r),
\end{equation}
where $\rho_{proj,0}(r,t/(c+1))$ is the projection of the spherical shell $\delta\big( r- {v t \over c+1} \big)/(4 \pi r^2)$ onto a diameter \cite{RGPN2012}. The key advantage provided by this procedure is that while the convolution in the second term takes place in 3 dimensions, the convolution in the third term takes place in 1 dimension. The function $\rho_{proj,0}$ is \cite{RGPN2012}:

\begin{equation}\label{roproj0}
\rho_{proj,0}(x,t) =
\left\{
\begin{array}{cc}
0, & x \notin [-vt, +vt]\\
{1/(2 v t)}, &\ \ \ \ \ \ x \in [-vt, +vt]\ \ \ \ \ \ \\
\end{array}
\right\},
\end{equation}
These are the sort of rectangle functions that need to be convoluted in one dimension.

So far we have summarized the procedure used in reference \cite{RGPN2012} to solve the Pearson random walk and {\it{formula \eqref{solution to Pearson} is valid only when collisions happen at regular intervals}}. But in the problem at hand collisions happen at uniformly distributed times. The pdf $\rho_c$ (defined by the expansion \eqref{PoissonExpansion}) can be found with the methods of reference \cite{RGPN2012} supplemented with two steps: first, we let the $c$ collisions happen at arbitrary times $t_1,..., t_c$; second, we average over all possible collision times $(t_1,..., t_c)$ (where $t_1<...<t_c)$, which amounts to a time convolution.

This approach can be summarized in the following diagram. The projection $\rho_{proj}$ of the exact solution is found by convoluting the projection $\rho_{proj,0}$, and then the projection is reversed by an inverse Abel transform:

$$
\begin{array}{ccc}
\ \ \rho\ \
{\underrightarrow{\ \ \ \ \ \ \ \ \ \ \text{\footnotesize{Projection onto a diameter}}\ \ \ \ \ \ \ \ \ \ } \atop \overleftarrow{\ \ \ \ \ \text{\footnotesize{Generalized inverse Abel transform}}\ \ \ \ \ }}
&\rho_{proj} \\
\end{array}
$$
\begin{equation}\label{esquema1}
\end{equation}

For $c=0$ the pdf $\rho_c$ is trivial, it is just the expanding spherical shell $\rho_0(r,t) = \delta(r-v t)/(4 \pi r^2)$. In the next subsection we find $\rho_c$ for $c=1$. A lot of work is required to find the $c=2$ solution and this is done in Appendix B.


\subsection{Exact solution for one collision}\label{Exact solution for one collision}

\subsubsection{Convolution of two rectangle functions}

We define the rectangle function:

\begin{equation}\label{}
{\rm{rect}}(w,x) \equiv
\left\{
\begin{array}{cc}
0, & x \in (-\infty, -w/2)\\
 & \\
{1 \over w}, &\ \ \ \ \ \ x \in [-w/2, +w/2]\ \ \ \ \ \ \\
 & \\
0, & x \in (+w/2, +\infty)
\end{array}
\right\},
\end{equation}
where $w$ is the width of the rectangle. This function is a pdf because it is positive and normalized. Moreover \cite{RGPN2012}

\begin{equation}\label{proj-rect}
\rho_{proj,0}(x,t) = {\rm{rect}}(2 vt,x),
\end{equation}
which is why we are interested in the rectangle function. The convolution of a function $f$ with a rectangle function is

$$
\big( {\rm{rect}}(w,\_) \otimes_s f \big)(x) = \int_{-\infty}^{+\infty}dx'\ f(x') {\rm{rect}}(w,x-x') =
$$

\begin{equation}\label{convrect}
\int_{x-w/2}^{x+w/2}dx'\ {f(x') \over w} = {1 \over w} \bigg( F\Big(x + {w \over 2}\Big) - F\Big(x - {w \over 2}\Big) \bigg),
\end{equation}
where $F$ is some primitive of $f$. Incidentally, this is known as the ``(simple) moving average'' in the social sciences.

The primitive of the rectangle function which we use is this continuous function

\begin{equation}\label{}
{\rm{Rect}}(w,x) \equiv
\left\{
\begin{array}{cc}
-{1 \over 2}, & x \in (-\infty, -w/2)\\
 & \\
\ \ {x \over w}, &\ \ \ \ \ \ x \in [-w/2, +w/2]\ \ \ \ \ \\
 & \\
+{1 \over 2}, & x \in (+w/2, +\infty)
\end{array}
\right\}.
\end{equation}
Then,

$$
\big( {\rm{rect}}(w_1,\_) \otimes_s {\rm{rect}}(w_2,\_) \big)(x) = {1 \over w_2} \bigg( {\rm{Rect}}\Big(w_1, x + {w_2 \over 2}\Big) - {\rm{Rect}}\Big(w_1, x - {w_2 \over 2}\Big) \bigg) =
$$

$$
\left\{
\begin{array}{cc}
\-{1 \over 2 w_2}, &x + {w_2 \over 2} \in (-\infty, -w_1/2)\\
 & \\
 \ {x + w_2/2 \over w_1 w_2}, &\ \ \ \ x + {w_2 \over 2} \in [-w_1/2, +w_1/2]\ \ \ \\
 & \\
 +{1 \over 2 w_2}, & x + {w_2 \over 2} \in (+w_1/2, +\infty)
\end{array}
\right\}
-
\left\{
\begin{array}{cc}
-{1 \over 2 w_2}, & x - {w_2 \over 2} \in (-\infty, -w_1/2)\\
 & \\
\ \ {x - w_2/2 \over w_1 w_2}, &\ \ \ \ x - {w_2 \over 2}\in [-w_1/2, +w_1/2]\ \ \ \\
 & \\
+{1 \over 2 w_2}, & x - {w_2 \over 2} \in (+w_1/2, +\infty)
\end{array}
\right\} =
$$

\begin{equation}\label{}
\left\{
\begin{array}{cc}
 -{1 \over 2 w_2}, & x \in \big( -\infty, {-w_1-w_2 \over 2} \big)\\
 & \\
 \ \ {x + w_2/2 \over w_1 w_2}, & \ \ \ \ x \in \big[ {-w_1 - w_2 \over 2}, {w_1 - w_2 \over 2} \big]\ \ \ \\
 & \\
 +{1 \over 2 w_2}, & x  \in \big( {w_1 - w_2 \over 2}, +\infty \big)
\end{array}
\right\}
-
\left\{
\begin{array}{cc}
 -{1 \over 2 w_2}, & x \in \big( -\infty, {-w_1+w_2 \over 2} \big)\\
 & \\
 \ \ {x - w_2/2 \over w_1 w_2}, & \ \ \ \ x \in \big[ {-w_1 + w_2 \over 2}, {w_1 + w_2 \over 2} \big]\ \ \ \\
 & \\
 +{1 \over 2 w_2}, & x  \in \big( {w_1 + w_2 \over 2}, +\infty \big)
\end{array}
\right\}.
\end{equation}

If we choose $w_1 < w_2$, then

$$
-\infty < {-w_1-w_2 \over 2} < {w_1-w_2 \over 2} < {-w_1+w_2 \over 2} < {+w_1+w_2 \over 2} < \infty
$$
and


\begin{equation}\label{conv2rect}
\big( \mathrm{rect}(w_1,\_) \otimes_s \mathrm{rect}(w_2,\_) \big)(x) =
\left\{ \begin{array}{cc}
0,&\displaystyle -\infty < x < {-w_1-w_2 \over 2} \\[12pt]
\displaystyle {1 \over 2}\Big({1 \over w_1} + {1 \over w_2}\Big) + {x \over w_1 w_2},
&\displaystyle \frac{-w_1-w_2}{2} \leq x \leq \frac{w_1-w_2}{2} \\[12pt]
\displaystyle \frac{1}{w_2},& \displaystyle \frac{w_1-w_2}{2} < x < \frac{-w_1+w_2}{2} \\[12pt]
\displaystyle \frac{1}{2}\Bigl(\frac{1}{w_1} + \frac{1}{w_2}\Bigr) - \frac{x}{w_1w_2},
&\displaystyle \frac{-w_1+w_2}{2} \leq x \leq \frac{w_1+w_2}{2} \\[12pt]
0, & \displaystyle \frac{w_1+w_2}{2} < x <\infty.
\end{array} \right\}
\end{equation}
The graph of $\big( {\rm{rect}}(w_1,\_) \otimes_s {\rm{rect}}(w_2,\_) \big)$ has the shape of a trapezoid whose base has length $w_1+w_2$ and whose top side has length $w_2-w_1$.

\subsubsection{Space-time convolution of two rectangle functions}

In order to present the results that follow it is convenient to define the function ``restriction'' by

\begin{equation}\label{}
{\rm{restr}}(a,x,b) \equiv
\left\{
\begin{array}{cc}
0, & x \in (-\infty, a)\\
1, &\ \ \ \ \ \ x \in [a, b]\ \ \ \ \ \ \\
0, & x \in (b, +\infty)
\end{array}
\right\}.
\end{equation}
Equation \eqref{conv2rect} requires $w_1 \leq w_2$. When $w \leq vt \Rightarrow w \leq 2 vt - w$, it follows from equation \eqref{conv2rect} that

$$
\big( {\rm{rect}}(w,\_) \otimes_s {\rm{rect}}(2 vt - w,\_) \big)(x) =
$$

$$
{\rm{restr}}(-\infty,x,-vt) \times 0 + {\rm{restr}}(-vt,x,w-vt) \Bigg( {1 \over 2} \l( {1 \over w} + {1 \over 2 vt-w} \r) + {x \over w (2 vt-w)} \Bigg) +
$$

$$
{\rm{restr}}(w-vt,x,-w+vt) {1 \over 2 vt-w} +
$$

\begin{equation}\label{conv2rect.1}
{\rm{restr}}(-w+vt,x,vt) \Bigg( {1 \over 2} \l( {1 \over w} + {1 \over 2 vt-w} \r) - {x \over w (2 vt-w)} \Bigg) + {\rm{restr}}(vt,x,+\infty) \times 0
\end{equation}
The equal sign means that the lhs and the rhs are equal in probability, that is, integrals over any interval have the same value for both of them. But the left-hand side and the right-hand side might differ at the end points of the intervals. This is irrelevant for our purposes.

The average over the collision times mentioned at the end of the second paragraph between eqs. \eqref{roproj0} and \eqref{esquema1}
is, for $c=1$,

$$
\rho_{proj,1}(x,t) = {1 \over t} \int_0^{t} dt_1\ \big( \rho_{proj,0}(\_,t_1) \otimes_s \rho_{proj,0}(\_,t-t_1) \big)(x) =
$$

\begin{equation}\label{TimeAv1}
{1 \over t} \int_0^{t} dt_1\ \big( {\rm{rect}}(2 v t_1,\_) \otimes_s {\rm{rect}}(2 v (t-t_1),\_) \big)(x).
\end{equation}
We use the fact that the above integrand is symmetric about $t_1 = t/2$ to rewrite $\rho_{proj,1}(x,t)$ as follows

\begin{equation}\label{}
\rho_{proj,1}(x,t) =
{2 \over t} \int_0^{t/2} dt_1\ \big( {\rm{rect}}(2 v t_1,\_) \otimes_s {\rm{rect}}(2 v (t-t_1),\_) \big)(x).
\end{equation}
With this new upper limit the inequality $2 v t_1 \leq vt$ holds and eq. \eqref{conv2rect.1} may be applied doing the substitution $w=2 v t_1$:

$$
\rho_{proj,1}(x,t) = {2 \over t} \int_0^{t/2} dt_1\ \Bigg[ {\rm{restr}}(-vt,x,2 v t_1-vt) \Bigg( {1 \over 2} \l( {1 \over 2 v t_1} + {1 \over 2 v (t-t_1)} \r) + {x \over 2 v t_1 (2 v (t-t_1))} \Bigg) +
$$

\begin{equation}\label{}
{{\rm{restr}}(2 v t_1-vt,x,-2 v t_1+vt) \over 2 v(t-t_1)} +
{\rm{restr}}(-2 v t_1+vt,x,vt) \Bigg( {1 \over 2} \l( {1 \over 2 v t_1} + {1 \over 2 v(t-t_1)} \r) - {x \over 2 v t_1 (2 v(t-t_1))} \Bigg) \Bigg].
\end{equation}
The integration limits imply the inequalities $0<t_1<t/2$. Simultaneously, $t_1$ has to satisfy the inequalities implied by the function restr. After solving for $t_1$ in these inequalities we obtain

$$
\rho_{proj,1}(x,t) = {2 H(-x) \over t} \int_{t/2+x/2v}^{t/2} dt_1\ \Bigg( {1 \over 2} \l( {1 \over 2 v t_1} + {1 \over 2 v (t-t_1)} \r) + {x \over 2 v t_1 (2 v (t-t_1))} \Bigg) +
$$

\begin{equation}\label{}
{2 \over t} \int_0^{t/2-|x|/2v} dt_1\ {1 \over 2 v(t-t_1)} +
{2 H(x)\over t} \int_{t/2-x/2v}^{t/2} dt_1\ \Bigg( {1 \over 2} \l( {1 \over 2 v t_1} + {1 \over 2 v (t-t_1)} \r) - {x \over 2 v t_1 (2 v (t-t_1))} \Bigg),
\end{equation}
where $H$ is a Heaviside function,

\begin{equation}\label{}
H(x) =
\left\{
\begin{array}{ccc}
  0 & {\rm{if}} & x < 0 \\
    &  &  \\
  1 & {\rm{if}} & x \geq 0
\end{array}
\right.
\end{equation}
The integrals can be done analytically to yield
\begin{equation}\label{}
\rho_{proj,1}(x,t) = {1 \over vt} \Bigg[ H(-x) {vt+x \over 2 vt} \ln {vt-x \over vt+x} + \ln {2 vt \over vt+|x|} + H(+x) {-vt+x \over 2 vt} \ln {vt-x \over vt+x} \Bigg].
\end{equation}
After some algebra, the preceding expression can be simplified to

\begin{equation}\label{rhor1}
\rho_{proj,1}(x,t) = {1 \over vt} \Bigg[ {vt+x \over 2 vt} \ln {vt-x \over vt+x} + \ln {2 vt \over vt-x} \Bigg],
\end{equation}
for all $x \in [-vt, +vt]$.

The three-dimensional pdf is found using the inverse Abel transform \eqref{inverseAbel.1}:

\begin{equation}\label{rho1}
\rho_1(\vec r,t) = -{1 \over 2 \pi r} {d \over dr} {1 \over vt} \Bigg[ {vt+r \over 2 vt} \ln {vt-r \over vt+r} + \ln {2 vt \over vt-r} \Bigg] = {1 \over 4 \pi (vt)^2 r} \ln {v t+r \over v t-r},
\end{equation}
which is graphed in Fig. \ref{Comparacionchoques.1}. Clearly, this function has a logarithmic divergence at $r=vt$. We believe that this has a qualitative consequence on the shape of the graph of the pdf, which is that $\rho$ (and also $\rho_I$ and $\rho_r$) has a vertical asymptote at $r=vt$. This issue is discussed in the conjecture of section \ref{Graphs} and commented upon in section \ref{Conclusions}.


When $\lambda t \ll 1$ (very little scattering and/or very early times)

\begin{equation}\label{}
\rho(\vec r, t) \approx {1 \over 1 + \lambda t} \l( {\delta(r-v t) \over 4 \pi r^2} + \lambda t \rho_1(\vec r, t) \r),
\end{equation}
which is the properly normalized sum of the expanding wave front and the first term of expansion \eqref{PoissonExpansion}.
\bigskip

\subsection{Exact solution for two and more collisions}\label{Exact solution for two and more collisions}

The exact solution for more than one collision could in principle be computed starting from the generalization of the average \eqref{TimeAv1} over the collision time $t_1$ to the case of $c$ collisions:

$$
\rho_{proj,c}(x,t) =
$$

$$
{c! \over t^c} \int_0^{t} dt_c\ \int_0^{t_c} dt_{c-1}\ \cdots \int_0^{t_2} dt_1\ \big( \rho_{proj,0}(\_,t_1) \otimes_s \rho_{proj,0}(\_,t_2-t_1) \otimes_s \cdots \otimes_s \rho_{proj,0}(\_,t-t_c) \big)(x) =
$$

\begin{equation}\label{conv.e-t}
{c! \over t^c} \int_0^{t} dt_c\ \int_0^{t_c} dt_{c-1}\ \cdots \int_0^{t_2} dt_1\ \rho_{proj,c}(t_1,...,t_c;x,t), \ \ {\rm{where}}\ \ t_1 < \cdots < t_c < t.
\end{equation}

This shows that $\rho_{proj,c}$ is the convolution over the time variable of rectangle functions of widths which add up to $vt$. Notice that the last equality defines the function $\rho_{proj,c}$. Taking the partial derivative of the last equality with respect to $x$ yields:

$$
{\partial \over \partial x} \rho_{proj,c}(x,t) =
$$

\begin{equation}\label{Dconv.e-t}
{c! \over t^c} \int_0^{t} dt_c\ \int_0^{t_c} dt_{c-1}\ \cdots \int_0^{t_2} dt_1\ {\partial \over \partial x} \rho_{proj,c}(t_1,...,t_c;x,t),
\end{equation}
which we are going to use in this subsection.
%

The derivation of the exact solution for two collisions is quite technical and does not incorporate any essential idea into the main body of the article. It can be safely skipped in a first reading and it is the content of Appendix B. The derivation of the exact solution for more than two collisions seems unfeasible.

However, two exact results can be derived for all the pdf conditional to $c$ collisions. In Appendix C (subsections \ref{Computation of rorc} and \ref{Computation of roIc}) these results will be used to estimate the error of the method presented in section \ref{Solutions as Fourier series}.

These results are:

{\bf{Proposition 5.1}}

\begin{equation}\label{Proposition 1}
{\rm{For}}\ c=1,2,3,...,\ \rho_{rs,c}(0) = \rho_{r,c}(0) = 0.
\end{equation}

{\bf{Proposition 5.2}}

\begin{equation}\label{Proposition 2}
{\rm{For}}\ c=2,3,4,...,\ \rho_{r,c}(vt) = \rho_{rs,c}(\pm vt) = \rho_{I,c}(vt) = \rho_{Is,c}(\pm vt) = 0.
\end{equation}

In order to prove each of these propositions we use two simple lemmas.

{\bf{Lemma 5.1}}. If two functions whose supports are contained in the intervals $[a,b]$ and $[c,d]$ are convoluted, the support of their convolution is contained in the interval $[a+c, b+d]$.

{\bf{Proof}}. It follows from the definition of convolution product. {\bf{QED}}.

{\bf{Lemma 5.2}}. Let a pdf $f$ satisfy $f \in C^r$, that is, its $r$-th derivative is continuous. Then its convolution with a rectangle function ${\rm{rect}}(w,\_)$ satisfies $f \otimes {\rm{rect}}(w,\_) \in C^{r+1}$.{\footnote{Using the definition of Riemann integral and the approximation $f(x) \approx \sum_i f(x_i + \Delta/2)$ restr$(x_i, x_i + \Delta)$, it is possible to prove that if $f_1 \in C^{r_1}$ and $f_2 \in C^{r_2}$, then $f_1 \otimes f_2 \in C^{{\rm{max}}(r_1,r_2)+1}.$}} Since $\rho_{proj,0}(x,t) = {\rm{rect}}(2 vt,x)$ (relation \eqref{proj-rect}), $\rho_{proj,0}(vt,\_) \in C^{-1}$ and this implies that the integrand of \eqref{conv.e-t} is in $C^{c-1}$, that is, $\rho_{proj,c}(t_1,...,t_c;\_,t) \in C^{c-1}$.

{\bf{Proof}}. It follows from formula \eqref{convrect}. {\bf{QED}}.

{\bf{Proof of Proposition 5.1}}. From the result \eqref{rhor1},

\begin{equation}\label{}
{\partial \over \partial x} \rho_{proj,1}(x,t) = {1 \over 2 v^2 t^2} \ln {vt-x \over vt+x},
\end{equation}
which is 0 at $x=0$. By the second of formulae \eqref{inverseAbel.1}, the proposition for $c=1$ is proven.

The symmetry $\rho_{proj,0}(-x,t) = \rho_{proj,0}(+x,t)$ is inherited by $\rho_{proj,c}(t_1,...,t_c;\_,t)$. Since the function $\rho_{proj,c}(t_1,...,t_c;\_,t)$ is symmetric and, from Lemma 5.2, it has a continuous derivative at $x=0$ for $c=2,3,4,...$, this derivative has to be 0. Proposition 5.1 follows from equality \eqref{Dconv.e-t} and the second of formulae \eqref{inverseAbel.1}. {\bf{QED}}.


{\bf{Proof of Proposition 5.2}}. By Lemma 5.1, ${\partial \over \partial x} \rho_{proj,c}(t_1,...,t_c;x,t) = 0$ when $x \notin [-vt, +vt]$. Therefore, ${\partial \over \partial x} \rho_{proj,c}(t_1,...,t_c;\pm vt,t) = 0$ if ${\partial \over \partial x} \rho_{proj,c}(t_1,...,t_c;\_,t)$ is continuous at $\pm vt$. Since, by Lemma 5.2, $\rho_{proj,c}(t_1,...,t_c;\_,t) \in C^{c-1}$, this happens for $c \geq 2$. Proposition 5.2 follows from equality \eqref{Dconv.e-t} and formulae \eqref{inverseAbel.1}. {\bf{QED}}.

\section{Solutions as Fourier series} \label{Solutions as Fourier series}

\subsection{Moments}\label{ssec:Moments}

To denote the moments we use the notation $\langle \ \rangle$ for expected value, as in Quantum Mechanics. The relations between the even moments of $\rho_{rs}(\_,t),\ \rho_{Is}(\_,t)$ and $\rho(\_,t)$ are (the odd moments will not be used):

\begin{equation}\label{Moments1}
\langle x^{2m} \rangle_{\rho_{rs}} = \langle x^{2m} \rangle_{\rho_{r}} = \langle r^{2m} \rangle_{\rho}\ \ {\rm{for}}\ m=0,1,2,...
\end{equation}
and

\begin{equation}\label{Moments2}
\langle x^{2m} \rangle_{\rho_{Is}} = \langle x^{2m} \rangle_{\rho_{I}} = {1 \over 4 \pi} \langle r^{2m-2} \rangle_{\rho} = {1 \over 4 \pi} \langle r^{2(m-1)} \rangle_{\rho}\ \ {\rm{for}}\ m=1,2,3,...,
\end{equation}
as follows from definitions \eqref{eq:definitions}-\eqref{rhoIs}.


For $m=0$ the last relation is still true, but $\langle r^{-2} \rangle_{\rho}$ cannot be obtained in the way in which the positive even moments are obtained in reference \cite{RGPN1995}. A more precise notation for the preceding moments would be $\langle x^{2m} \rangle_{\rho_{rs}(\_,t)}$, etc, but we have preferred to keep the notation light. Bear in mind that all these moments depend on time.
\bigskip

For the pdf $\rho_S$ \eqref{decomposition} there is a Poisson expansion analogous to expansion \eqref{PoissonExpansion} which is

\begin{equation}\label{PoissonExpansion.s}
\rho_S(\vec r, t) = e^{-\lambda t} \sum_{c=0}^\infty {(\lambda t)^c \over c!} \rho_{S,c}(\vec r, t).
\end{equation}
The $\rho_{S,c}$'s of this expansion are the pdf's conditional to $c$ collisions and are normalized, just like the $\rho_{c}$'s of expansion \eqref{PoissonExpansion}, so they are the same. The skeptical reader can check this statement using the identity $e^{-\lambda t} = (1 - e^{-\lambda t})/(e^{\lambda t}-1)$ and formulae \eqref{decomposition},  \eqref{PoissonExpansion} and \eqref{PoissonExpansion.s}.

Either from formulae (25) and (26) of reference \cite{RGPN1995} or from formulae (31) and (32) of reference \cite{RGPN2012cap} for the 3-dimensional case we obtain:

\begin{equation}\label{EvenMoment}
\langle r^{2 m} \rangle_{\rho_S} =
(2 m+1)!\ e^{- \lambda t}\ (vt)^{2 m} \sum_{c=0}^\infty {(\lambda t)^c \over (2 m + c)!}\ C(c,m),
\end{equation}
where

\begin{equation}\label{C(c,m)def}
C(c,m) \equiv \sum_{\underset{i_1 + \cdots + i_{c+1} = m} {i_1, \cdots, i_{c+1} \in N}} {1 \over (2 i_1+1) \cdots (2 i_{c+1}+1)}
\end{equation}
and $c$ labels the number of collisions. See Appendix A for the computation and some properties of the $C(c,m)$ coefficients.

It follows from formula \eqref{EvenMoment} and from collision expansion \eqref{PoissonExpansion} that

$$
\langle x^{2m} \rangle_{\rho_{rs,c}} = \langle x^{2m} \rangle_{\rho_{r,c}} = \langle r^{2 m} \rangle_{\rho_c} = \langle r^{2 m} \rangle_{\rho_{S,c}} =
$$

\begin{equation}\label{EvenMomentrc}
(vt)^{2 m} {c! (2 m+1)! \over (2 m + c)!}\ C(c,m)\ \ {\rm{for}}\ m=0,1,2,...
\end{equation}

It follows from formulae \eqref{Moments2} and \eqref{EvenMomentrc} that

\begin{equation}\label{EvenMoment1c}
\langle x^{2 m} \rangle_{\rho_{Is,c}} =
{(vt)^{2 m-2} \over 4 \pi} {c! (2 m-1)! \over (2 m -2 + c)!}\ C(c,m-1)\ \ {\rm{for}}\ m=1,2,3,...
\end{equation}
These very simple time dependences are a consequence of the scaling relation \eqref{ScalingRhoc1}.

\subsection{Fourier expansion in terms of the moments}

The definition of Fourier transform that we use is
\begin{equation}\label{}
\tilde f(\nu) \equiv \int dx\ f(x)\ e^{-i 2 \pi \nu x} = \langle e^{-i 2 \pi \nu x} \rangle = \sum_{m=0}^\infty {(-i 2 \pi)^m \langle x^m\rangle \over m!} \nu^m,
\end{equation}
where the last equality holds when the support of $f$ is bounded, as in the pdf's of interest in this article. To show this, notice that in our problem $\langle x^m\rangle \leq (vt)^m$ so that the above series is bounded, term by term, by the series expansion of $\exp(2 \pi v t \nu)$, which proves convergence.

For any periodic function $f_p$ of period $2vt$ (see the beginning of section 3 of reference \cite{RGPN2023}),
\begin{equation}\label{}
f_p(x) =
{1 \over 2vt} \sum_{h=-\infty}^{+\infty} \tilde f \l( {h \over 2vt} \r) e^{i {2 \pi \over 2vt} h x}.
\end{equation}
It follows from the two previous equations that, when $f_p$ is even,

\begin{equation}\label{ExpansionEvenFunction}
f_p(x) =
{1 \over 2vt} \l( \langle x^{0} \rangle + 2 \sum_{h=1}^{\infty} \left[ \sum_{m=0}^{\infty} {(- 4 \pi^2)^m \langle x^{2m} \rangle \over (2m)!} \cdot \l( {h \over 2vt} \r)^{2m} \right] \cos {\pi h x \over vt} \r).
\end{equation}

A word of caution on the notation. We have chosen to work with the symmetrized pdf's (the ones that have $s$ as subindex) in order to be able to use the above expansion, which has only even moments. So the index $m$ refers not to the $m$-th moment, but to the $2m$-th moment.

\section{Fourier series of $\rho_{rs,c}$ and $\rho_{Is,c}$}\label{Fourier series of r and I}

We are going to find $\rho(x,t)$ in three dimensions by means of a Fourier series. The Fourier-Laplace transform of $\rho$ is known \cite{Claes1987, Stadje1989}, but it cannot be inverted analytically. But the moments of $\rho(\_, t)$ are known \cite{RGPN1995}, and from them a Fourier series for $\rho(\_, t)$ can be obtained as in one dimension \cite{RGPN2023}. Indeed, if we consider $\rho_{rs,c}(\_,t): [-vt,+vt] \rightarrow \Re^+$ as the restriction to $[-vt,+vt]$ of a periodic function of period $2 vt$, then we may expand it using a Fourier series. Substitution of expression \eqref{EvenMomentrc} into formula \eqref{ExpansionEvenFunction} yields, for $x \in [-vt, +vt]$,

\begin{equation}\label{rorsc}
\rho_{rs,c}(x,t) = {1 \over 2 v t}\ \l( 1 + 2 \sum_{h=1}^{\infty} \left[\sum_{m=0}^\infty (-\pi^2)^m {(2m+1) c! \over (2m+c)!} C(c,m) \cdot h^{2m} \right] \cos {\pi h \over v t} x \r).
\end{equation}
Likewise, substitution of expressions \eqref{Moments2} (for $\langle r^0 \rangle_{\rho_{Is,c}}$) and \eqref{EvenMoment1c} (for $\langle r^{2m} \rangle_{\rho_{Is,c}},\ m=1,2,3,...$)  into formula \eqref{ExpansionEvenFunction} yields, for $x \in [-vt, +vt]$,

$$
\rho_{Is,c}(x,t) =
$$

\begin{equation}\label{ro1sc.1}
{1 \over 2 v t}\ \l( {\langle r^{-2} \rangle_{\rho_c}(t) \over 4 \pi} + {1 \over 2 \pi} \sum_{h=1}^{\infty} \left[ \langle r^{-2} \rangle_{\rho_c}(t) + \sum_{m=1}^\infty { (-\pi^2)^m c!\ C(c,m-1) \over 2 m (vt)^2 (2m-2+c)!} \cdot h^{2m} \right] \cos {\pi h \over v t} x \r).
\end{equation}
Note that, unlike formula \eqref{rorsc}, in the previous formula the lower index of the summation sign over $m$ starts at 1, because the contribution of $\langle r^0 \rangle_{\rho_{Is,c}}$ has been pulled out of the summation sign. In formula \eqref{rorsc} one can check that the dependence on time of $\rho_{rs,c}(x,t)$ is the one stated in equation \eqref{ScalingRhoc1}. The scaling relation \eqref{ScalingRhoc2} applied to $\rho_{Is,c}(x,t)$ implies

\begin{equation}\label{ScalingMenos2}
\langle r^{-2} \rangle_{\rho_c}(t) = {\langle r^{-2} \rangle_{\rho_c}(1/v) \over (vt)^2}.
\end{equation}

Theoretically, we could find $\rho_{Is,c}$ from the equation

\begin{equation}\label{rI}
\rho_{rs,c}(x,t) = 4 \pi x^2 \rho_{Is,c}(x,t)
\end{equation}
and avoid using expansion \eqref{ro1sc.1}, in which the quantity $\langle r^{-2} \rangle_{\rho_c}(t)$ is not provided by reference \cite{RGPN1995}. Note that equality \eqref{rI} requires, for $\rho_{Is,c}(x,t)$ not to diverge at $x=0$, that $\lim_{x \to 0} (\rho_{rs,c}(x,t)/x^2)$ is a constant. But in expression \eqref{rorsc} the upper limits of the summation signs cannot actually be made infinity. Thus $\rho_{rs,c}(x,t)$ cannot be computed exactly around $x=0$ and ${\rho_{rs,c}(x,t)/(4 \pi x^2)}$ is bound to be numerically not well behaved near the origin, as shown in Fig. \ref{Comparacionchoques.1}.

But there is an elegant and accurate way of finding $\langle r^{-2}\rangle_{\rho}(t)$ in expression \eqref{ro1sc.1}. When $vt = 1$ the Fourier series coefficients of expansion \eqref{ro1sc.1} are

\begin{equation}\label{FSro1.0}
FS(\rho_{Is,c})(0) \equiv
{\langle r^{-2} \rangle_{\rho_c}(1/v)  \over 8 \pi}
\end{equation}
and

\begin{equation}\label{FSro1}
FS(\rho_{Is,c})(h) \equiv
{1 \over 4 \pi} \left[ \langle r^{-2} \rangle_{\rho_c}(1/v) + \sum_{m=1}^\infty { (-\pi^2)^m c!\ C(c,m-1) \over 2 m (2m-2+c)!} \cdot h^{2m} \right] \ \ \ {\rm{for}}\ \ \ h=1,2,3,...
\end{equation}
When $vt \neq1$ the pdf $\rho_{Is,c}$ may be computed using the scaling relations \eqref{ScalingRhoc1}:

$$
\rho_{Is}(x, t) = {1 \over e^{\lambda t}-1} \sum_{c=1}^\infty {(\lambda t)^c \over c!} \rho_{Is,c}(x, t) = {1 \over e^{\lambda t}-1} \sum_{c=1}^\infty {(\lambda t)^c \over c!} {1 \over (vt)^3} \rho_{Is,c}\Big( {x \over vt}, {1 \over v} \Big) =
$$

\begin{equation}\label{Desarrollo}
{1 \over e^{\lambda t}-1} {1 \over (vt)^3} \sum_{c=1}^\infty {(\lambda t)^c \over c!} \sum_{h=0}^\infty FS(\rho_{Is,c})(h) \cos {\pi h \over v t} x.
\end{equation}

We now show that $\lim_{h \to \infty} FS(\rho_{Is,c})(h) = 0,\ \ \forall c \in \left\{ 1,2,3,... \right\}$, which will allow us to solve for $\langle r^{-2} \rangle_{\rho_c}(1/v)$ in eq. \eqref{FSro1}. Intuitively, if $FS(\rho_{Is,c})(h)$ didn't tend to zero as $h \rightarrow \infty$, then there would be oscillations of unbounded frequency and $\rho_{Is,c}$ would then not be a continuous function. We now give a proof of the advertised limit.

{\bf{Proposition 7.1.}} Let there be a natural number $n$. Let there be a function $f:\Re^n \rightarrow \Re$ which has minimum $m$ and maximum $M$. Let there be a pdf $g:\Re^n \rightarrow \Re^+$.
Then the minimum $m'$ and maximum $M'$ of their convolution $f \otimes g$ satisfy $[m',M'] \subset [m,M]$.

{\bf{Proof.}} The expected value of $f$ with respect to any pdf, which we denote by $\langle f \rangle$, certainly satisfies $\langle f \rangle \in [m, M]$. The convolution $f \otimes g$ at $\vec r \in \Re^n$ is

$$
(f \otimes g)(\vec r) = \int d^nr'\ f(\vec r') g(\vec r - \vec r') = \int d^nr'\ f(\vec r') g(-(\vec r' - \vec r)),
$$
which is the expected value of $f$ with respect to the pdf $- T_{\vec r}(g) \equiv g(-(\_ - \vec r))$, that is, the pdf $g$ translated by the vector $\vec r$ and then inverted. Therefore $(f \otimes g)(\vec r) \in [m, M]$, which proves the proposition. {\bf{QED}}.

{\bf{Proposition 7.2.}} If $\rho_c$ is bounded by $M_c \in \Re^+$, then $\rho_{c+1}$ is bounded by $(c+1) M_{c}$.

{\bf{Proof.}} It follows from formula \eqref{roc} that

\begin{equation}\label{convProp2}
\rho_{c+1}(\vec r, t) = {c+1 \over t} \big( \rho_{c} \otimes \rho_{0}\big) (\vec r, t) = {c+1 \over t} \int_0^t dt'\ \big( \rho_{c}(\_,t') \otimes_s \rho_{0}(\_,t-t') \big) (\vec r').
\end{equation}
Proposition 7.1 implies that the integrand in the last term is bounded by $M_c$ for all $t' \in [0,t]$. The integrand is then averaged over $t'$ and multiplied by $c+1$, thus the final result is bounded by $(c+1) M_c$. {\bf{QED}}.

{\bf{Proposition 7.3.}} $\lim_{h \to \infty} FS(\rho_{Is,c})(h) = 0,\ \ \forall c \in \left\{ 1,2,3,... \right\}$.

{\bf{Proof.}} By Parseval's formula (see, e. g., reference \cite{Champeney1987}), the $L^2$ norm squared of $\rho_{Is,c}$ is
\begin{equation}\label{}
||\rho_{Is,c}||_2^2 = {1 \over vt} \bigg( FS(\rho_{Is,c})(0)^2 + {1 \over 2} \sum_{h=0}^\infty FS(\rho_{Is,c})(h)^2 \bigg).
\end{equation}
Suppose that $\lim_{h \to \infty} FS(\rho_{Is,c})(h) \neq 0$. Then $||\rho_{Is,c}||_2^2$ would be infinite. But we can see that
\begin{equation}\label{}
||\rho_{Is,c}||_2 < \infty\ \ \ {\rm{for}}\ c=1,2,3,...
\end{equation}
This can be checked directly for $\rho_{Is,1}$ and $\rho_{Is,2}$, for which we have derived analytic expressions. Indeed, $\rho_{Is,1}$ can be found straightforwardly from the result \eqref{rho1} and definitions \eqref{rhoI} and \eqref{rhoIs}, bearing in mind expansion \eqref{PoissonExpansion}. Likewise, $\rho_{Is,2}$ can be found straightforwardly from the result for $\rho_{proj,2}$ stated right before definition \eqref{rho2} and definition \eqref{rho2} itself, and definitions \eqref{rhoI} and \eqref{rhoIs}, bearing in mind expansion \eqref{PoissonExpansion}.

But $\rho_{Is,2}$ is not only of bounded $L_2$ norm, it is also bounded, as can be seen in the graph of $\rho_{Is,2}(r,t)$ in Fig. \ref{Comparacionchoques.1}. By applying Proposition 7.2 iteratively to $\rho_2$, $\rho_3$, $\rho_4$,..., it follows that the $\rho_c$'s are bounded for $c \geq 2$, which implies that their $L^2$ norm is finite. {\bf{QED}}.


We define a quantity related to $FS(\rho_{Is,c})(h)$:

\begin{equation}\label{FS_0}
FS_0(\rho_{Is,c})(h) \equiv {1 \over 4 \pi} \sum_{m=1}^\infty { (-\pi^2)^m c!\ C(c,m-1) \over 2 m (2m-2+c)!} \cdot h^{2m}\ \ \ {\rm{for}}\ \ \ h=1,2,3,....
\end{equation}
The result $\lim_{h \to \infty} FS(\rho_{Is,c})(h) = 0,\ \ \forall c \in \left\{ 1,2,3,... \right\}$ and eq. \eqref{FSro1} imply that

\begin{equation}\label{limFS}
\langle r^{-2} \rangle_{\rho_c}(1/v) = - \lim_{h \to \infty} 4 \pi\ FS_0(\rho_{Is,c})(h),\ \ \forall c \in \left\{ 1,2,3,... \right\}
\end{equation}
which, using scaling \eqref{ScalingMenos2}, renders formula \eqref{ro1sc.1} useful. This limit is evaluated numerically for each $c$ using expression \eqref{FS_0}. An example of this (for $c=3$) is worked out in section \ref{Computation of roIc} of Appendix C.

\section{Graphs and simulations}\label{Graphs}

In \cite{RGPNgooglesite} there are four Mathematica files which produce a function called Ro[r, t], which is $(1 - e^{-\lambda t}) \rho_I(r,t)$, the pdf of the scattered particles. They have to be compiled in the order 1,2,3,4. The values for the speed $v$ of the particle and the scattering rate $\lambda$ need to be provided by the user. The function Ro[r, t] is $(1 - e^{-\lambda t}) \rho_I(r,t)$, where $\rho_I$ is twice the series \eqref{Desarrollo} when $\lambda t < 100$; the Gaussian approximation presented in subsection ``Asymptotics'' of Appendix C when $\lambda t > 105$; and the linear interpolation between the two previous cases when $100 \leq \lambda t \leq 105$. Though readers without access to the program Mathematica will be unable to produce graphs for themselves, we hope that - being Mathematica a high level programming language - these readers will still understand the code. In Fig. \ref{Final graphs roI} we show graphs of $(1 - e^{-\lambda t}) \rho_I(r,t)$ for $\lambda t=$0.1, 1 and 10.

\begin{figure}[h]
\centering
\includegraphics[width=0.33 \textwidth]{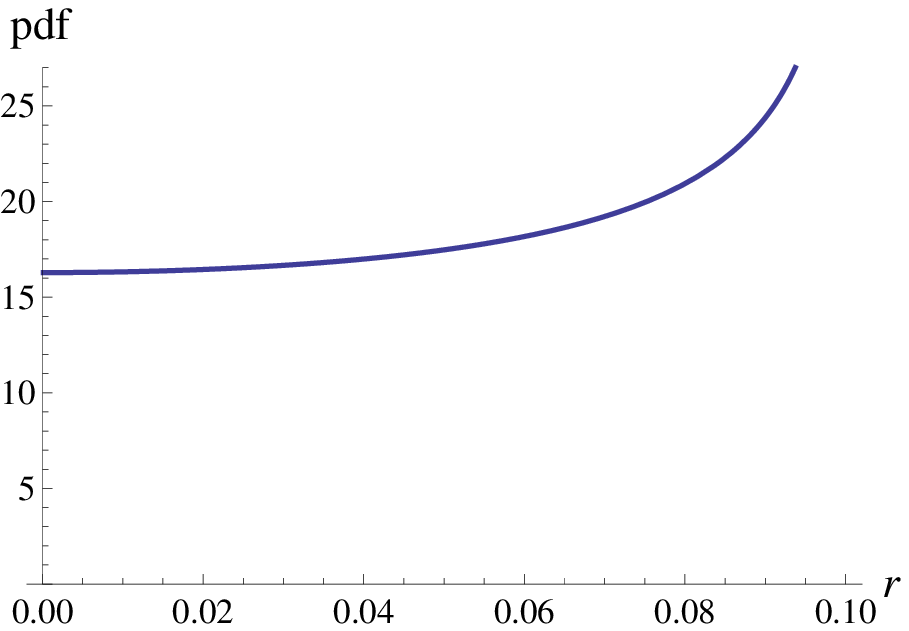}\includegraphics[width=0.33 \textwidth]{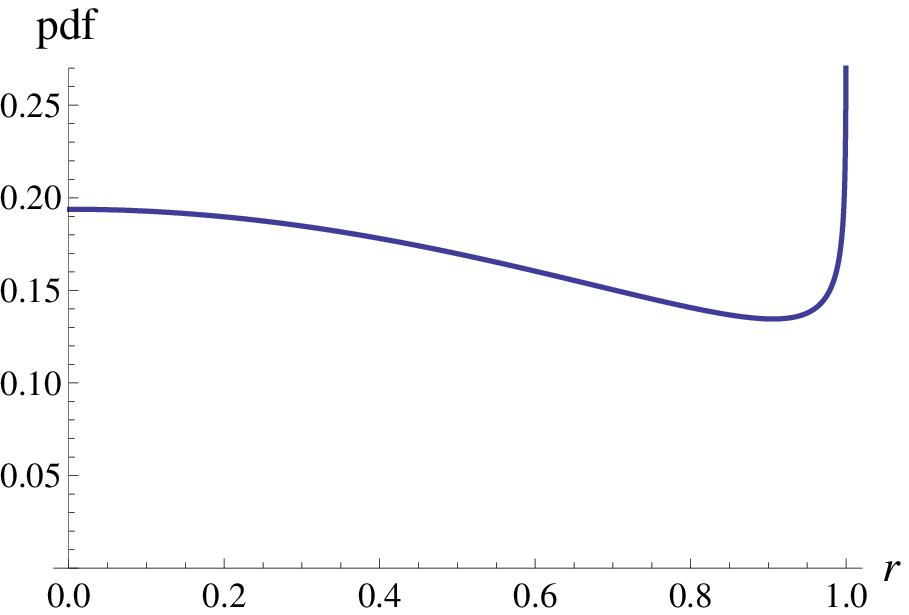}\includegraphics[width=0.33 \textwidth]{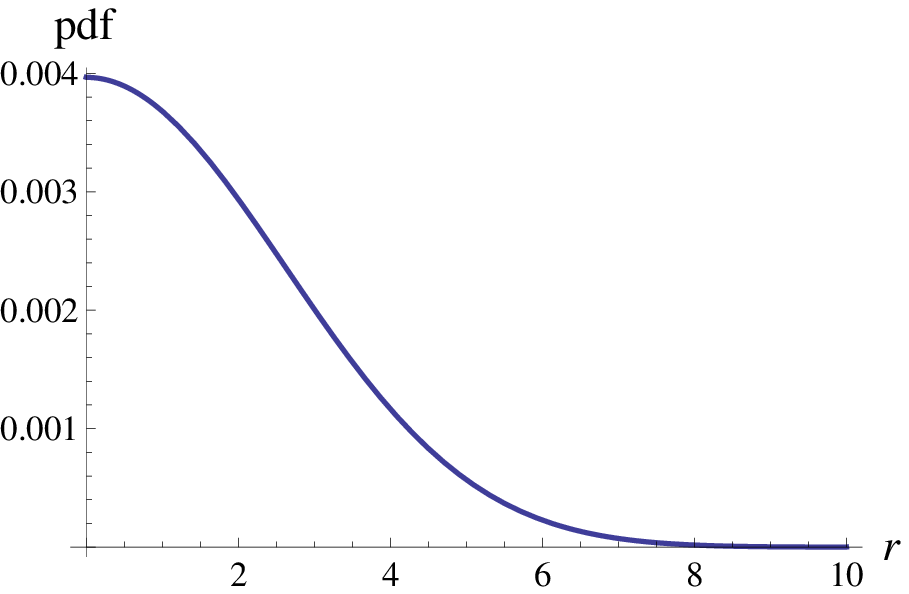}
\caption{Graphs of $(1 - e^{-\lambda t}) \rho_I(r,t)$ for $\lambda = 1$, $v=1$ and (from left to right) $t=0.1,1,10$. Notice the different scales in each graph.}\label{Final graphs roI}
\end{figure}

Graphical representations of $(1 - e^{-\lambda t}) \rho_r(r,t) = 4 \pi r^2 (1 - e^{-\lambda t}) \rho_I(r,t)$, which is the probability density at $r$ when the integration is done over $dr$ (as opposed to integrating over $4 \pi r^2 dr$), are shown in Fig. \ref{Final graphs ror}.
\begin{figure}[h]
\centering
\includegraphics[width=0.33 \textwidth]{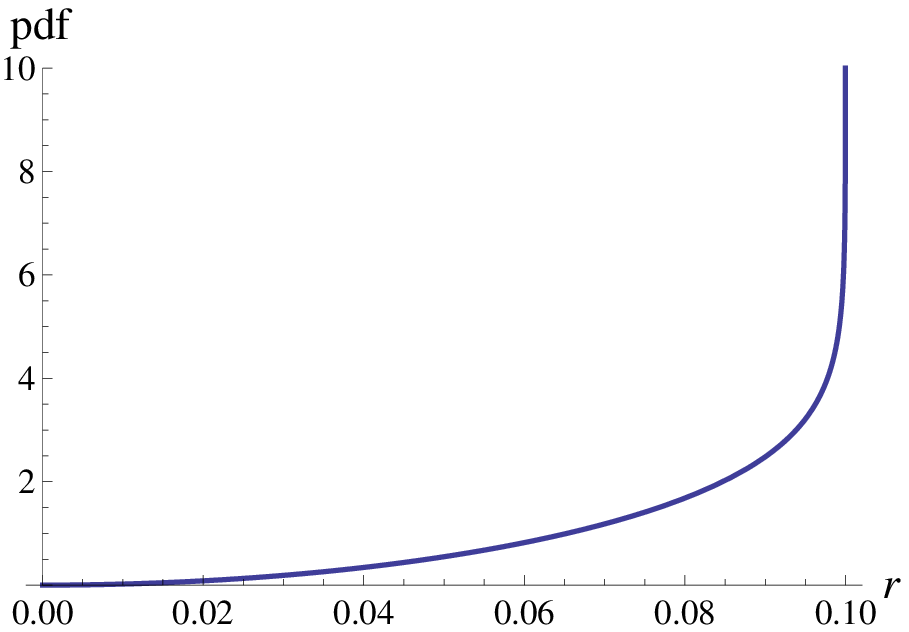}\includegraphics[width=0.33 \textwidth]{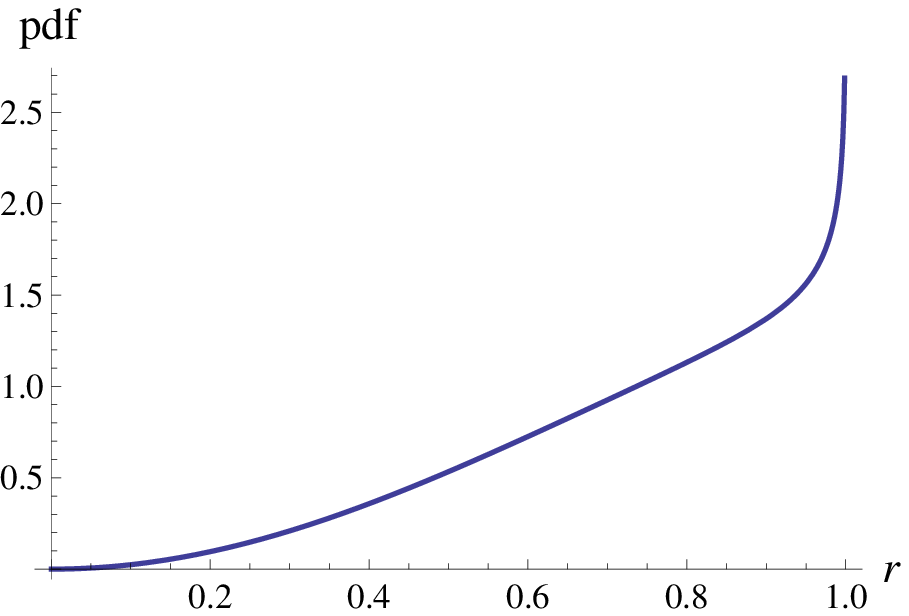}\includegraphics[width=0.33 \textwidth]{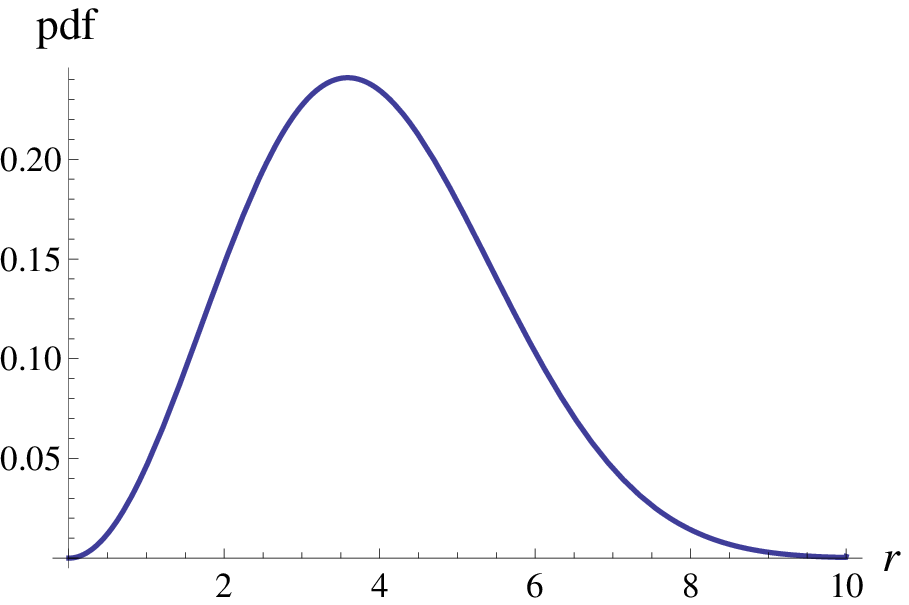}
\caption{Graphs of $(1 - e^{-\lambda t}) \rho_r(r,t)$ for $\lambda = 1$, $v=1$ and (from left to right) $t=0.1,1,10$. Notice the different scales in each graph.}\label{Final graphs ror}
\end{figure}

The pdf $\rho_1$ \eqref{rho1} has a logarithmic divergence at $vt$ shown in Fig. \ref{Comparacionchoques.1}. Of course $\rho_{r,1}$ also has a logarithmic divergence at $vt$. Figures \ref{Final graphs roI}, \ref{Final graphs ror} and \ref{Final graphs divergences} and Proposition 5.2 strongly suggest the following

{\bf{Conjecture.}}

\begin{equation}\label{conjecture}
\lim_{r \to vt} { \lambda t \rho_{I,1}(r,t) \over (e^{\lambda t} - 1) \rho_{I}(r,t) } = \lim_{r \to vt} { \lambda t \rho_{r,1}(r,t) \over (e^{\lambda t} - 1) \rho_{r}(r,t) } = 1.
\end{equation}
This conjecture needs proof because it is conceivable that for $c \geq 2$ the $\rho_c$'s add up in the vicinity of $r=vt$ to yield a divergence which modifies or even cancels the one due to $\rho_1$.

\begin{figure}[h]
\centering
\includegraphics[width=0.33 \textwidth]{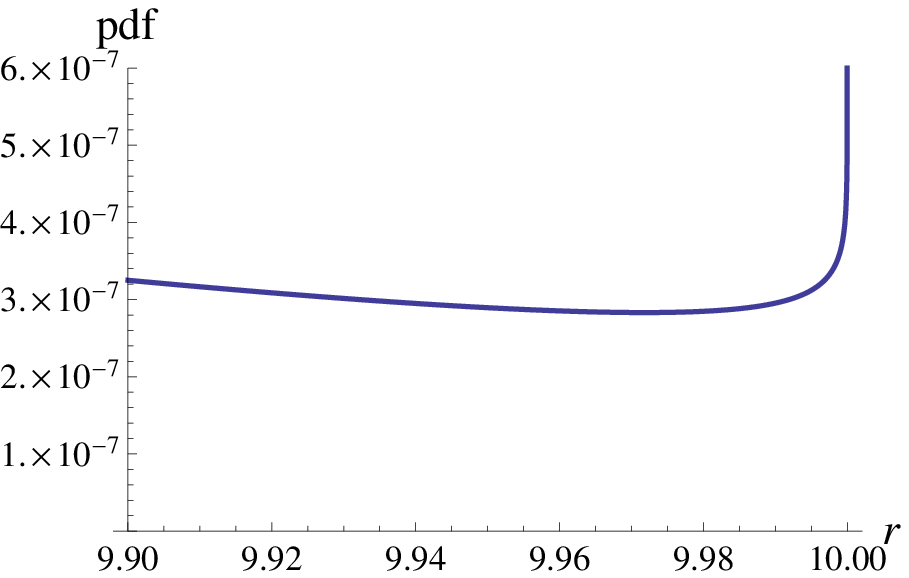}\hspace{1cm}\includegraphics[width=0.33 \textwidth]{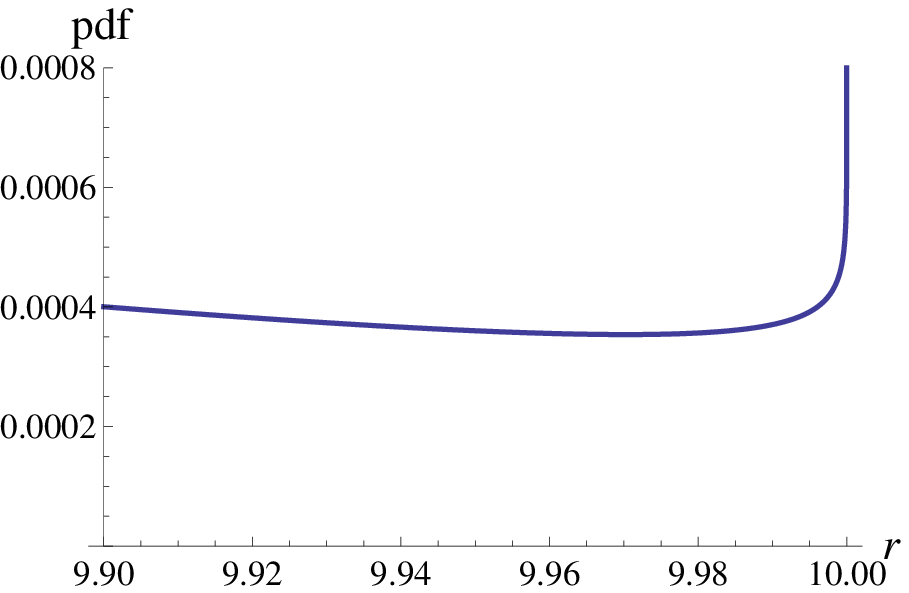}
\caption{The conjectured divergences at $r=vt$ of $(1 - e^{-\lambda t}) \rho_I(r,t)$ and $(1 - e^{-\lambda t}) \rho_r(r,t)$ for $\lambda = 1$, $v=1$ and $t=10$ become visible at the right scale }\label{Final graphs divergences}
\end{figure}

The correctness of the results presented in this article is firmly supported by the match shown in Figs. \ref{Comparacion1choque} and \ref{Comparacion2choques} in Appendix C between the two very different methods presented in sections \ref{Exact solution} and \ref{Solutions as Fourier series}. Nevertheless, for the benefit of the reader who is
interested in simulations for their own sake, in the webpage \cite{RGPNgooglesite} there is a Mathematica file called Simulations.nb in which the essential ingredient to simulate the random flight is given. We have simulated random flights with the condition that they change direction thrice and compared the resulting distribution of position with $\rho_{I,3} = 2 \rho_{Is,3}$, as computed in subsection \ref{Computation of roIc}. When the number of random flights reaches 10,000, the distributions obtained from the simulations and from $\rho_{I,3}$ are undistinguishable from each other (Fig. \ref{Simulations}). Since the validity of the collision expansion \eqref{PoissonExpansion} is beyond question, the match of the comparison between the simulations and our results for any number $c$ of collisions implies the match of a complete comparison.

\begin{figure}[h]
\centering
\includegraphics[width=0.45 \textwidth]{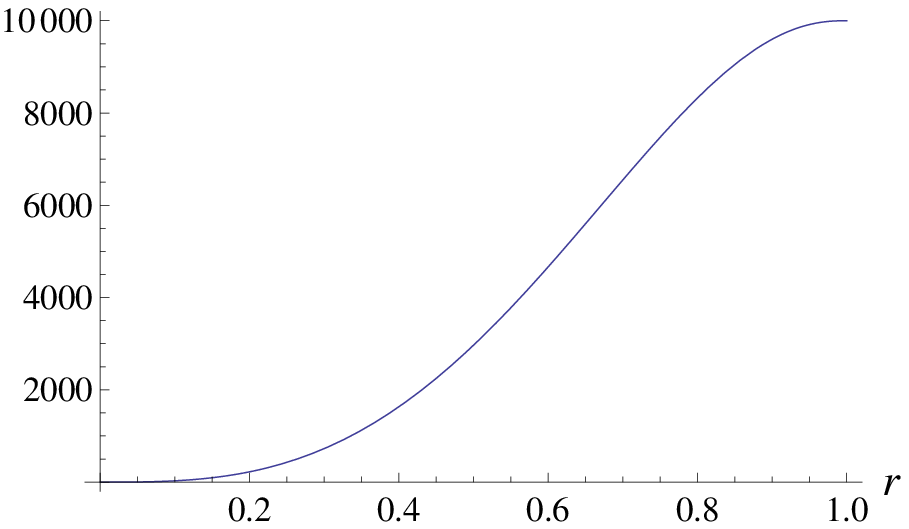}\hspace{1cm}\includegraphics[width=0.45 \textwidth]{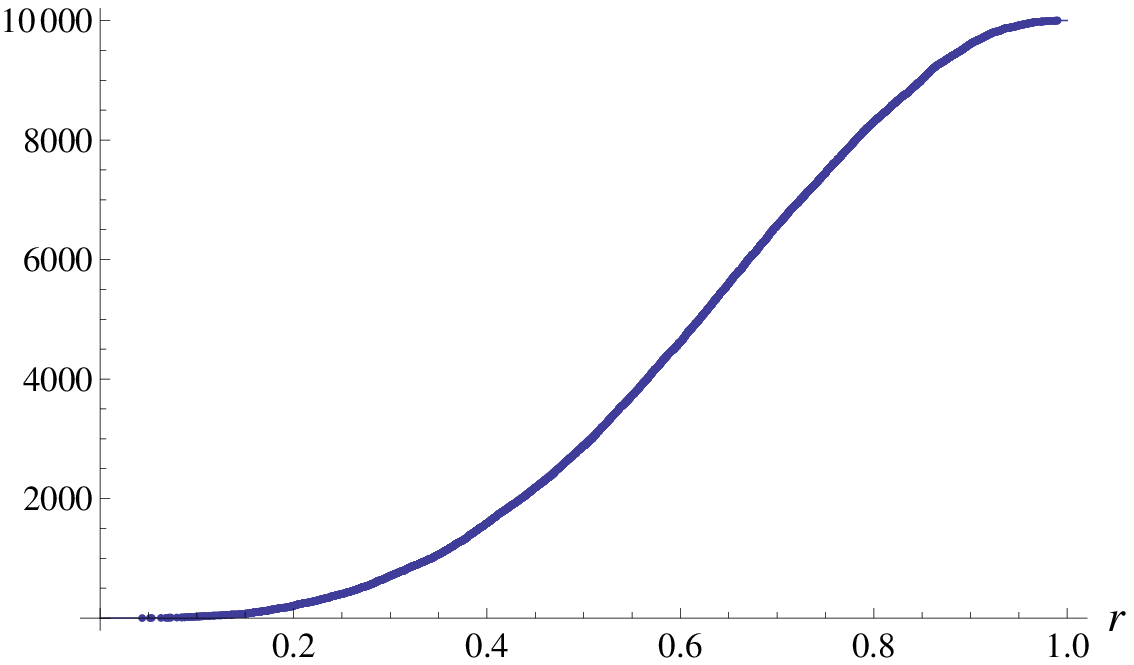}
\caption{In the left figure the cumulative function $\int_0^r dr' \rho_{r,3}(r',1/v)$ multiplied by $10^4$ is depicted. In the right figure the graph of the right figure is plotted again and, in addition, there are 10,000 dots, each corresponding to a simulation. In it each dot has as ordinate the number of simulations whose length is shorter than its abscissa. One can see that the dots fall right on the theoretical curve.}\label{Simulations}
\end{figure}

\section{Conclusions}\label{Conclusions}

The two distinct lines of attack of the present article are summarized in the diagram which follows. It is just a summary. In particular, it glosses over the fact that the supports of each of the three pdf's shown in the diagram are not the same. Indeed, the supports of $\rho_{rs}$ and $\rho_{proj}$ are $[-vt, +vt]$, but the support of $\rho$ is the sphere of radius $vt$. In the approach presented in section \ref{Exact solution} we start with an exact solution for $\rho_{proj}$ and use the generalized inverse Abel transform to undo the projection. This is the approach shown on the right side of the diagram, which is based on the results of reference \cite{RGPN2012}. It is analytically feasible only for 1 and 2 collisions. On the left hand side, the Fourier series approach, which was presented in \cite{RGPN2023}, and which is feasible for any number of collisions. 

$$
\begin{array}{ccc}
{\huge{\rho_{rs}}}&
{\underrightarrow{\ \ \ \ \ \ \ \ \ \ \ \ \ \ \ \ \ $$1 \over 2 \pi x^2$$\ \ \ \ \ \ \ \ \ \ \ \ \ \ \ \ \ } \atop \overleftarrow{\ \ \ \ \ \ \ \ \ \ \ \ \ \ \ \ \ $$\scriptstyle{2 \pi r^2}$$\ \ \ \ \ \ \ \ \ \ \ \ \ \ \ \ \ }}
\ \ \rho\ \
{\underleftarrow{\text{\footnotesize{Generalized inverse Abel transform}}} \atop \overrightarrow{\ \ \ \ \ \text{\footnotesize{Projection onto a diameter}}\ \ \ \ \ }}
&\rho_{proj} \\
\uparrow & & \uparrow \\
\end{array}
$$
{\footnotesize{Fourier series}}\hspace{4in}{\footnotesize{\ \ \ \ \ Exact solution from}}\\
{\footnotesize{from the moments}}\hspace{3.7in}{\footnotesize{space-time convolution}}
\begin{equation}\label{esquema2}
\end{equation}


Conjecture \eqref{conjecture} is important because, if true, it provides a simple analytic form for $\rho$ in the wake of the expanding front. Furthermore, suppose that we want to know the density of scatterers using a pulse of light. This simple analytic form allows to solve for $\lambda$ easily even if the intensity of the light source is unknown.

We have used a method already presented in reference \cite{RGPN2012} and a newer method presented in reference \cite{RGPN2023} and in section \ref{Solutions as Fourier series} of this article. This newer method is essentially how to retrieve a pdf from its moments. While widely known that this can be done under a wide set of conditions (see, e. g., section 6.4 of the book by Moran \cite{Moran1968}), how to actually compute the pdf from its moments is still a subject of research \cite{Tagliani2003,Mnatsakanov2008,Novi2023} which can be applied to a number of fields \cite{RGPN2011,RGPN2013, Stoyanov2016}.

The results of this article allow to solve this random flight with arbitrary initial conditions, not just the isotropic one. The topic of initial conditions in the random flight has been discussed seldom \cite{Stadje1989,RGPN1998,Kolesnik2008,RGPN2012}. Here we are going to follow the approach of section 5 of reference \cite{RGPN2012cap}.
There it is shown that if the solution to the ``bullet initial condition'' \cite{RGPN2012cap} is known, then the problem with arbitrary initial conditions can be solved. In the bullet initial condition the probability density of the yet unscattered particles is $e^{-\lambda t} \delta(x-v t, y, z)$, where we have taken as the positive direction of the $x$-axis the direction along which the unscattered particles travel. The solution $\rho_b$ to the bullet initial condition can be put in terms of the solution to the isotropic initial condition as follows \cite{RGPN1998}{\footnote{In the integral which follows $\rho_S$ takes the same values as the $\rho_S$ of formula \eqref{decomposition.0}, but it is not the same function. This is because the domain of the function $\rho_S$ of formula \eqref{decomposition.0} is $B_{(\vec 0, vt)} \times \Re^+$, whereas the domain of the function $\rho_S$ of the above integral is $[0, vt] \times \Re^+$. The relation between both functions is the same as the functions related by definition \eqref{rhoI}. Since this issue arises only once, we decided not to introduce new notation.}}:

\begin{equation}
\rho_b (\vec r,t) = e^{-\lambda t} \delta(v t - x) \delta(y) \delta(z) + \lambda \int_0^t dt'\ e^{-\lambda t'} \rho_S(|\vec r - v t' \vec i|, t-t').
\end{equation}
The above integral has been done analytically in the two-dimensional case \cite{RGPN1998,Stadje1989}, but this seems impossible or quite difficult in 3 dimensions, where we think that  numerical integration would be necessary.
\bigskip

{\Large{\bf{Data Availability Statement}}}

The data that supports the findings of this study are available in the supplementary material of this article
\bigskip

{\Large{\bf{Acknowledgements}}}

This work was supported by MINECO/AEI and FEDER/EU under Project PID2020-112576GB-C21. The author thanks the MINECO/AEI of Spain for the financial support.

\section{Appendix A. The coefficients $C(c,m)$}\label{}

\bigskip

The coefficients $C(c,m)$ are defined by

\begin{equation}\label{}
C(c,m) \equiv \sum_{\underset{i_1 + \cdots + i_{c+1} = m} {i_1, \cdots, i_{c+1} \in N}} {1 \over (2 i_1+1) \cdots (2 i_{c+1}+1)}.
\end{equation}
They can be computed using nested summations, that is

\begin{equation}\label{}
C(c,m) =
\sum_{i_1=0}^m {1 \over 2 i_1+1} \sum_{i_2=0}^{m-i_1} {1 \over 2 i_2+1} \cdots \sum_{i_{c}=0}^{m-i_1-\cdots-i_{c-1}} {1 \over (2 i_c+1)} {1 \over 2 (m-i_1-\cdots-i_c)+1}.
\end{equation}
With symbolic software it might be more convenient to compute $C(c,m)$ as follows. We remark that $C(c,m)$ is the coefficient of $x^m$ in $\l( 1 + {x \over 3} + {x^2 \over 5} + \cdots \r)^{c+1}$. This series is reminiscent of the expansion of $\ln (1+x)$. Indeed,

$$
1 + {x \over 3} + {x^2 \over 5} + \cdots = f(x) \equiv {1 \over 2 \sqrt{x}} \ln {1 + \sqrt{x} \over 1 - \sqrt{x}}\ \ ({\rm{for}}\ 0<x<1).
$$
Thus,

\begin{equation}\label{}
C(c,m) = {1 \over m!} \left.{d^m \over dx^m}\right|_{x=0} \l( f(x) \r)^{c+1}.
\end{equation}
However, the computation is faster if one defines the $m$-th degree polynomial

\begin{equation}\label{}
f(m;x) \equiv \sum_{i=0}^m {x^i \over 2 i + 1}
\end{equation}
and then
\begin{equation}\label{ftruncada}
C(c,m) = {1 \over m!} \left.{d^m \over dx^m}\right|_{x=0} \l( f(m;x) \r)^{c+1}.
\end{equation}

It can be read off from the definition that

\begin{equation}\label{}
C(0,m) = {1 \over 2m +1}.
\end{equation}

Bounds for $C(c,m)$ can be obtained as follows. The number of terms in the sum
\begin{equation}\label{}
C(c,m) \equiv \sum_{\underset{i_1 + \cdots + i_{c+1} = m} {i_1, \cdots, i_{c+1} \in N}} {1 \over (2 i_1+1) \cdots (2 i_{c+1}+1)}
\end{equation}
is the number of ways in which $c$ separators may be placed between $m$ objects, which is ${m+c \choose m}$ (see, e. g., \cite{Feller1966I}, chapter 2, section 5). The largest term is $1 \over 2 m + 1$, and $\Big({1 \over 2 (m/(c+1)) + 1}\Big)^{c+1}$ is a lower bound for all the terms. Therefore,

\begin{equation}\label{}
{m+c \choose m} \bigg({1 \over 2 (m/(c+1)) + 1}\bigg)^{c+1} \leq C(c,m) \leq {m+c \choose m} {1 \over 2 m + 1}.
\end{equation}
\medskip

The first few $C(c,m)$ coefficients (up to the 8th moment and 5 collisions) are in Table 1:
\medskip

\begin{tabular}{|c||c|c|c|c|c|}
\hline
\multicolumn{6} { | c | }{Table 1: The first few $C(c,m)$ coefficients.}\\
\hline
  \hline
  c $\backslash$ m & 0 & 1 & 2 & 3 & 4 \\
  \hline \hline
  0 & 1 & 1/3 & 1/5 & 1/7 & 1/9 \\
  \hline
  1 & 1 & 2/3 & 0.511 & 0.419 & 0.357 \\
  \hline
  2 & 1 & 3/3 & 0.933 & 0.866 & 0.806 \\
  \hline
  3 & 1 & 4/3 & 1.467 & 1.520 & 1.535 \\
  \hline
  4 & 1 & 5/3 & 2.111 & 2.418 & 2.636 \\
  \hline
  5 & 1 & 6/3 & 2.867 & 3.598 & 4.214 \\
  \hline
\end{tabular}


\bigskip

\section{Appendix B: Exact solution for two collisions}\label{}

\subsection{Convolution of three rectangle functions}

In order to obtain the convolution of three rectangle functions we are going to decompose the trapezoid which is the convolution of two rectangle functions into a central rectangle and one triangle on each side. Two collisions happening at times $t_1, t_2$ split the time $t$ into the segments $[0, t_1]$, $[t_1, t_2]$ and $[t_2, t]$, and the length of the total run into three segments of lengths $v t_1$, $v(t_2-t_1)$ and $v(t-t_2)$. For the purposes of this Appendix, it will be convenient to denote by $a/2$, $b/2$ and $c/2$ the lengths, in increasing order of length (that is, not respectively), of these three segments. Let then $a, b$ and $c$ be positive real numbers such that $a + b + c = 2 vt$ and $0<a<b<c<2 vt$.

Since $2 vt -b-c \leq b \leq c$,

$$
{\rm{rect}}(2 vt - b - c,\_) \otimes_s \big( {\rm{rect}}(b,\_) \otimes_s {\rm{rect}}(c,\_) \big)(x) =
$$

$$
{\rm{rect}}(2 vt - b - c,\_) \otimes_s \Bigg( {\rm{restr}} \Big({-b-c \over 2},\_,{-c+b \over 2} \Big) \bigg( {1 \over 2} \Big( {1 \over b} + {1 \over c} \Big) + {\_ \over b c} \bigg) +
$$

\begin{equation}\label{conv3rect}
{\rm{restr}} \Big( {-c+b \over 2},\_,{c-b \over 2} \Big) {1 \over c} + {\rm{restr}} \Big({c-b \over 2},\_,{b+c \over 2} \Big) \bigg( {1 \over 2} \Big( {1 \over b} + {1 \over c} \Big) - {\_ \over b c} \bigg) \Bigg)(x).
\end{equation}
All of the terms which appear when the preceding expression is developed are proportional to the already seen convolution of two rectangle functions except for the terms ${\rm{rect}}(2 vt - b - c,\_) \otimes_s \Big( {\rm{restr}}({-b-c \over 2},\_,{-c+b \over 2}) {\_ \over b c} \Big)(x)$ and $-{\rm{rect}}(2 vt - b - c,\_) \otimes_s \Big( {\rm{restr}}({c-b \over 2},\_,{b+c \over 2}) {\_ \over b c} \Big)(x)$. According to formula \eqref{convrect}, the first of the new terms is

$$
{\rm{rect}}(2 vt - b - c,\_) \otimes_s \Big( {\rm{restr}}\Big({-b-c \over 2},\_,{-c+b \over 2} \Big) {\_ \over b c} \Big)(x) =
$$

\begin{equation}\label{}
{1 \over 2 vt - b - c} \int_{x-vt+{b+c \over 2}}^{x+vt-{b+c \over 2}} dx'\ {\rm{restr}}\Big({-b-c \over 2},x',{-c+b \over 2} \Big) {x' \over b c}.
\end{equation}
The $x$ which appears in the limits of the integral in the second line cannot take any values, because $x$ has to satisfy $x-vt+{b+c \over 2} \leq {b-c \over 2}$ and $x+vt-{b+c \over 2} \geq -{b+c \over 2}$. The ${\rm{restr}}(-vt,x,vt-c)$ prefactor which follows enforces the satisfaction of both inequalities.

$$
{\rm{rect}}(2 vt - b - c,\_) \otimes_s \Big( {\rm{restr}}\Big({-b-c \over 2},\_,{-c+b \over 2} \Big) {\_ \over b c} \Big)(x) =
$$

$$
{ {\rm{restr}}(-vt,x,vt-c) \over 2bc (2 vt - b - c)} \l( \min\Big(x+vt-{b+c \over 2},{-c+b \over 2}\Big)^2 - \max\Big(x-vt+{b+c \over 2},{-b-c \over 2}\Big)^2 \r) =
$$

$$
{ 1 \over 2bc (2 vt - b - c)} \Bigg[
{\rm{restr}}(-vt,x,vt-b-c) \l( \Big(x+vt-{b+c \over 2}\Big)^2 - \Big({-b-c \over 2}\Big)^2 \r) +
$$

$$
{\rm{restr}}(vt-b-c,x,-vt+b) \l( \Big(x+vt-{b+c \over 2}\Big)^2 - \Big(x-vt+{b+c \over 2}\Big)^2 \r) +
$$

$$
{\rm{restr}}(-vt+b,x,vt-c) \l( \Big({-c+b \over 2}\Big)^2 - \Big(x-vt+{b+c \over 2}\Big)^2 \r) \Bigg] =
$$

$$
-{ 1 \over 2bc (2 vt - b - c)} \Bigg[
{\rm{restr}}(-vt,x,vt-b-c) \Big( (b + c - vt - x) (vt + x) \Big) +
$$

$$
{\rm{restr}}(vt-b-c,x,-vt+b) \Big( 2 (b + c - 2 vt) x \Big) +
$$

\begin{equation}\label{conv2rect.2}
{\rm{restr}}(-vt+b,x,vt-c) \Big( (b - vt + x) (c - vt + x) \Big) \Bigg].
\end{equation}
This function is quadratic and concave up (i. e., as $x^2$) in the first interval, linear in the second interval and quadratic and concave down (i. e., as $-x^2$) in the third interval.

The other new convolution product, $-{\rm{rect}}(2 vt - b - c,\_) \otimes_s \Big( {\rm{restr}}({c-b \over 2},\_,{b+c \over 2}) {\_ \over b c} \Big)(x)$, is quadratic and concave down  in the first interval, linear in the second interval and quadratic and concave up in the third interval.

%
%

Formulae \eqref{conv2rect} and \eqref{conv2rect.2} give us the two kinds of convolution products that we need to do the convolution of three rectangle functions. Now we simply apply the distributive property of the convolution product with respect to the sum. Before we do that it is convenient to recast results \eqref{convrect} and \eqref{conv2rect}. Since

\begin{equation}\label{lastthree1}
{\rm{rect}}(w,x) = {1 \over w} {\rm{restr}} \Big(-{w \over 2},x,+{w \over 2} \Big)\ \Rightarrow\ {\rm{restr}} (a, x, b) = (b-a) {\rm{rect}} \Big( b-a, x-{a+b \over 2} \Big),
\end{equation}
similarly to formula \eqref{convrect} we obtain

\begin{equation}\label{lastthree2}
\big( {\rm{restr}}(a,\_,b) \otimes_s f \big)(x) = F(x - a) - F(x - b),
\end{equation}
where $F$ is some primitive of $f$, and similarly to \eqref{conv2rect} we obtain

$$
\big( {\rm{restr}}(x_1,\_,x_2) \otimes_s {\rm{restr}}(x_3,\_,x_4) \big)(x) =
$$

\begin{equation}\label{lastthree3}
\left\{
\begin{array}{cc}
0, & x \in \big( -\infty, x_3+x_1 \big)\\
+x-x_3-x_1, &\ \ \ \ x \in \big[ x_3+x_1, x_3+x_2 \big]\ \ \ \\
x_2-x_1, & x  \in \big( x_3+x_2, x_4+x_1 \big) \\
-x+x_2+x_4, & x  \in \big[ x_4+x_1, x_4+x_2 \big] \\
0, & x \in \big( x_4+x_2, +\infty \big)\\
\end{array}
\right\},
\end{equation}
provided $x_2-x_1 \leq x_4-x_3 \Leftrightarrow x_3+x_2\leq x_4+x_1$.

The last three formulae (\ref{lastthree1}-\ref{lastthree3}) serve to expand the convolution \eqref{conv3rect} of three rectangle functions. Since $c$ is the side of the longest rectangle, ${2 vt \over 3} \leq c$ is always satisfied. The three functions involved in the convolution product \eqref{conv3rect} are proportional to restr$\big( vt - {b+c \over 2}, -vt + {b+c \over 2} \big)$, restr$\big( - {b \over 2}, + {b \over 2} \big)$ and restr$\big( - {c \over 2}, + {c \over 2} \big)$. There are eight possible sums of the three limits of these restr functions, taking one limit from each function. These sums are the limits between the different regions of the piecewise defined convolution product, similarly to the convolution product \eqref{lastthree3}. One can check that when ${2 vt \over 3} \leq c \leq vt$, the ordered sums are

\begin{equation}\label{}
-vt,-b-c+vt,b-vt,c-vt,-c+vt,-b+vt,b+c-vt\ {\rm{and}}\ vt,
\end{equation}
and that when $vt \leq c \leq 2 vt$, the ordered sums are

\begin{equation}\label{}
-vt,-b-c+vt,b-vt,-c+vt,c-vt,-b+vt,b+c-vt\ {\rm{and}}\ vt.
\end{equation}
These remarks make the following results more transparent.
\medskip

In the case ${2 vt \over 3} \leq c \leq vt$:

$$
f_1(vt,c,b,x) \equiv
{\rm{rect}}(2 vt - b - c,\_) \otimes_s \big( {\rm{rect}}(b,\_) \otimes_s {\rm{rect}}(c,\_) \big)(x) =
$$

$$
{1 \over 2 vt-b-c} \Bigg[ 0 \times {\rm{restr}}(-\infty,x,-vt) +
{(vt + x)^2 \over 2bc} {\rm{restr}}(-vt,x,-b-c+vt) +
$$

$$
{(b + c + 2 x) (2 vt-b-c) \over 2bc} {\rm{restr}}(-b-c+vt,x,b-vt) -
$$

$$
{2 b^2 + 2 b (c - 2 vt) + (c-vt+x)^2 \over 2bc} {\rm{restr}}(b-vt,x,c-vt) -
$$

$$
{b^2 + b (c - 2 vt) + (c - vt)^2 + x^2 \over bc} {\rm{restr}}(c-vt,x,-c+vt) -
$$

$$
{2 b^2 + 2 b (c - 2 vt) + (c-vt-x)^2 \over 2bc} {\rm{restr}}(-c+vt,x,-b+vt) +
$$

$$
{(b + c - 2 x) (2 vt-b-c) \over 2bc} {\rm{restr}}(-b+vt,x,b+c-vt) +
$$

\begin{equation}\label{f1}
{(vt - x)^2 \over 2bc} {\rm{restr}}(b+c-vt,x,vt) +
0 \times {\rm{restr}}(vt,x,\infty) \Bigg].
\end{equation}

In the case $vt \leq c \leq 2 vt$:

$$
f_2(vt,c,b,x) \equiv
{\rm{rect}}(2 vt - b - c,\_) \otimes_s \big( {\rm{rect}}(b,\_) \otimes_s {\rm{rect}}(c,\_) \big)(x) =
$$

$$
{1 \over 2 vt-b-c} \Bigg[ 0 \times {\rm{restr}}(-\infty,x,-vt) +
{(vt + x)^2 \over 2bc} {\rm{restr}}(-vt,x,-b-c+vt) +
$$

$$
{(b + c + 2 x) (2 vt-b-c) \over 2bc} {\rm{restr}}(-b-c+vt,x,b-vt) -
$$

$$
{2 b^2 + 2 b (c - 2 vt) + (c-vt+x)^2 \over 2bc} {\rm{restr}}(b-vt,x,-c+vt) +
$$

$$
{2 vt - b - c \over c} {\rm{restr}}(-c+vt,x,c-vt) -
$$

$$
{2 b^2 + 2 b (c - 2 vt) + (c-vt-x)^2 \over 2bc} {\rm{restr}}(c-vt,x,-b+vt) +
$$

$$
{(b + c - 2 x) (2 vt-b-c) \over 2bc} {\rm{restr}}(-b+vt,x,b+c-vt) +
$$

\begin{equation}\label{f2}
{(vt - x)^2 \over 2bc} {\rm{restr}}(b+c-vt,x,vt) +
0 \times {\rm{restr}}(vt,x,\infty) \Bigg].
\end{equation}

For later use in the three next figures (\ref{fig:tri-x-grandes}, \ref{fig:tri-x-medianos} and \ref{fig:tri-x-pequenos}) we assign notation to the different analytic forms which appear in the convolution of three rectangle functions:

\begin{equation}\label{h}
\left.
\begin{array}{cc}
h_1(vt,c,b,x) \equiv & {(vt - |x|)^2 \over 2 b c (2 vt - b - c)} \\
\\
h_2(vt,c,b,x) \equiv & {(b + c - 2 |x|) \over 2bc} \\
\\
h_3(vt,c,b,x) \equiv & - {2 b^2 + 2 b (c - 2 vt) + (c-vt-|x|)^2 \over 2bc (2 vt-b-c)} \\
\\
h_4(vt,c,b,x) \equiv & - {b^2 + b (c - 2 vt) + (c-vt)^2 + x^2 \over bc (2 vt-b-c)} \\
\\
h_{4'}(vt,c,b,x) \equiv & {1 \over c} \\
\end{array}
\right\}
\end{equation}

\subsection{Space-time convolution of three rectangle functions}


We define

\begin{equation}\label{}
f(vt,c,b,x) \equiv
\left\{
\begin{array}{cc}
f_1(vt,c,b,x), & {2 vt \over 3} \leq c \leq vt \\
f_2(vt,c,b,x), & vt \leq c \leq 2 vt \\
\end{array}
\right\}.
\end{equation}
In order to find $\rho_{proj,2}$ we may average $f_1$ and $f_2$ as in integral \eqref{conv.e-t}. However, we have found it more convenient to integrate over the variables $a, b$ and $c$ defined in the first paragraph of this section. Since $\int_{2 vt \over 3}^{vt}dc\ \int_{2 vt - c \over 2}^{c}db\ 1 + \int_{vt}^{2 vt}dc\ \int_{2 vt - c \over 2}^{2 vt - c}db\ 1 = {v^2 t^2 \over 3}$, the wanted space-time convolution is the average

$$
{3 \over v^2 t^2} \int_{2 vt \over 3}^{2 vt}dc\ \int_{2 vt - c \over 2}^{\min(c,2 vt - c)}db\ f(vt,c,b,x) =
$$

\begin{equation}\label{}
{3 \over v^2 t^2} \l( \int_{2 vt \over 3}^{vt}dc\ \int_{2 vt - c \over 2}^{c}db\ f_1(vt,c,b,x) +
\int_{vt}^{2 vt}dc\ \int_{2 vt - c \over 2}^{2 vt - c}db\ f_2(vt,c,b,x) \r).
\end{equation}

The above integrals have a different structure depending on whether

\begin{equation}\label{}
\left.
\begin{array}{ccc}
  |x| & \in & [0,vt/3] \\
  |x| & \in & [vt/3,vt/2] \\
  |x| & \in & [vt/2,vt]
\end{array}
\right\}.
\end{equation}
These structures will be developed now and a figure depicting a triangle will show each of the three cases. For simplicity, the prefactor ${3 \over v^2 t^2}$ will be omitted.

In each of the three figures an isosceles right triangle is depicted for which $b+c < 2 v t$. This right triangle is divided into 6 triangles. The region of integration is the upper right triangle. This triangle is the set of points for which $b<c$ and ${2 v t - c \over 2} < b$, that is, the set of points for which $a < b < c$. For each of the three cases that follow, there are 4 different integration regions within the said triangle, separated by thin, continuous lines. In each of these four regions the function $h_1, h_2, h_3, h_4$ or $h_{4'}$ which needs to be integrated is written. The space-time convolution that we want to compute has the $x \leftrightarrow -x$ symmetry. Because of this symmetry we need to compute the said convolution only for $x < 0$.
\medskip

The four integrals in the case $|x| \in [vt/2,vt]$, when $x < 0$, are

\newpage

\begin{figure}[h]
\centering
\includegraphics[width=0.99 \textwidth]{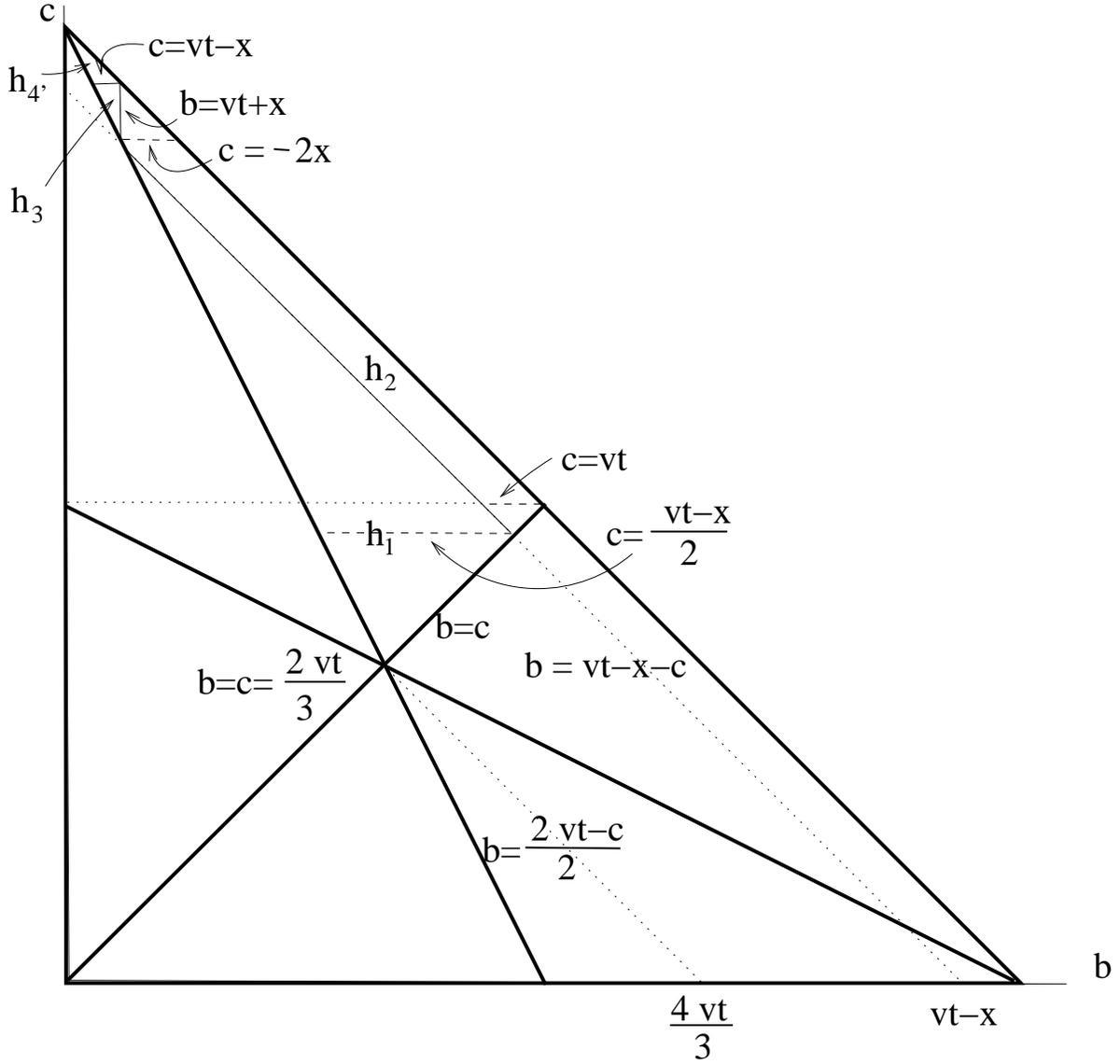}
\caption{\label{fig:tri-x-grandes}The above tiling showing the different integrands and regions of integration in the $(b,c)$ plane is valid for $-vt \leq x \leq -vt/2$.}
\end{figure}

$$
{\rm{restr}}(-vt,x,-vt/2)
$$

$$
\Bigg(
\int_{2 vt \over 3}^{vt-x \over 2}dc\ \int_{2 vt - c \over 2}^{c}db\ {(vt + x)^2 \over 2 b c (2 vt - b - c)} +
\int_{vt-x \over 2}^{-2 x}dc\ \int_{2 vt - c \over 2}^{vt-c-x}db\ {(vt + x)^2 \over 2 b c (2 vt - b - c)}
\Bigg) =
$$

$$
{(vt+x)^2 \over 8 vt} {\rm{restr}}(-vt,x,-vt/2)
$$

$$
\Bigg( \pi^2 - 3 (\ln2)^2 - \ln 3 \ln {243 \over 63} + \big(\ln vt\big)^2 - \Li_2\bigg( {1 \over 9} \bigg) - 6 \Li_2\bigg( {2 \over 3} \bigg) + \big(\ln(vt - x)\big)^2 +
$$

$$
2 \Bigg[ \bigg( \arg \tanh {vt \over 2 vt+x} - \arg \tanh {x \over vt} \bigg) \ln 4 + \ln {vt+x \over vt-x} \ln(3 vt + x) - \ln(vt + x) \ln vt
$$

\begin{equation}\label{roproj2.1}
- \Li_2\bigg( {vt - x \over 4 vt} \bigg)
+ \Li_2\bigg( {vt - x \over 2 vt} \bigg)
+\Li_2\bigg( -{vt + x \over 2 vt} \bigg)
+\Li_2\bigg( {vt + x \over vt - x} \bigg)
- \Re\bigg( \Li_2\bigg( {1 \over 2} + {vt \over vt + x} \bigg) \bigg) \Bigg] \Bigg),
\end{equation}
where $\Li_2$ is the dilogarithm function.

$$
{\rm{restr}}(-vt,x,-vt/2)
$$

$$
\Bigg(
\int_{vt - x \over 2}^{vt}dc\ \int_{vt - c - x}^{c}db\ {b + c + 2 x \over 2 b c} +
\int_{vt}^{-2 x}dc\ \int_{vt - c - x}^{2 vt-c}db\ {b + c + 2 x \over 2 b c} +
$$

$$
\int_{-2 x}^{vt-x}dc\ \int_{vt + x}^{2 vt-c}db\ {b + c + 2 x \over 2 b c}
\Bigg) =
$$

$$
{ {\rm{restr}}(-vt,x,-vt/2) \over 2}
$$

$$
\Bigg(
-vt - x - x (\ln 2)^2 - x \ln vt - x \ln 4 \ln vt - x (\ln vt)^2 +
  vt \ln(vt - x) - x \ln(vt - x) + x \ln 4 \ln(vt - x) +
$$

$$
x (\ln(vt - x))^2 + x \ln(-(vt/x)) + x \ln(-8 x) + x \ln(-2 x) +
x \ln 4 \ln(-2 x) + 2 x \ln vt \ln(-2 x) -
$$

$$
2 x \ln(vt - x) \ln(-2 x) + x \ln(-x) - vt \ln(vt + x) -
x \ln(4 (vt + x)) + 2 x \ln(-2 x) \ln {vt + x \over 2 vt}
$$

\begin{equation}\label{}
-2 x \ln(vt - x) \ln {vt + x \over vt} -
2 x \Li_2 \Big( {vt - x \over 2 vt} \Big) + 2 x \Li_2 \Big( { -2 x \over vt - x} \Big)
\Bigg).
\end{equation}

$$
{\rm{restr}}(-vt,x,-vt/2) \int_{-2x}^{vt-x}dc\ \int_{2 vt - c \over 2}^{vt + x}db\ {2 b^2 + 2 b (c - 2 vt) + (c-vt+x)^2 \over 2bc (b+c-2 vt)} =
$$

\begin{equation}\label{}
{ {\rm{restr}}(-vt,x,-vt/2) \over vt} \Bigg( vt (vt + x) - {\pi^2 \over 48} (vt + x)^2 + vt x \ln {vt - x \over - 2 x} -
{1 \over 4} (vt - x)^2 \Li_2 {vt + x \over vt - x} \Bigg).
\end{equation}

$$
{\rm{restr}}(-vt,x,-vt/2) \int_{vt-x}^{2 vt}dc\ \int_{2 vt - c \over 2}^{2 vt - c}db\ {1 \over c} =
$$

\begin{equation}\label{}
{\rm{restr}}(-vt,x,-vt/2)\Bigg( -{vt + x \over 2} + vt \ln {2 vt \over vt - x} \Bigg).
\end{equation}

\newpage

The four integrals in the case $|x| \in [vt/3,vt/2]$, when $x < 0$, are

\begin{figure}[h]
\centering
\includegraphics[width=0.99 \textwidth]{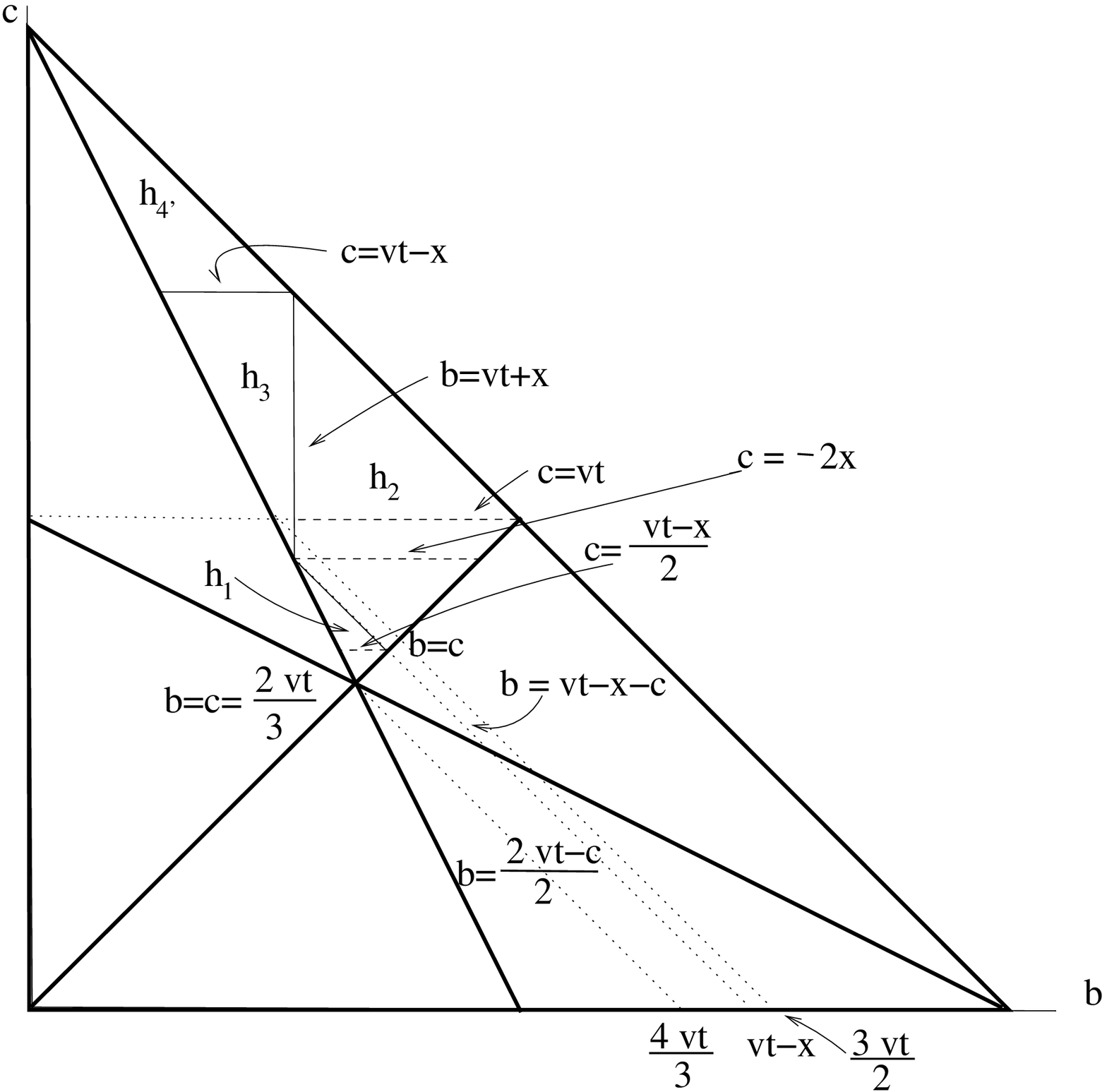}
\caption{\label{fig:tri-x-medianos}The above tiling showing the different integrands and regions of integration in the $(b,c)$ plane is valid for $-vt/2 \leq x \leq -vt/3$.}
\end{figure}

$$
{\rm{restr}}(-vt/2,x,-vt/3)
$$

$$
\Bigg(
\int_{2 vt \over 3}^{vt-x \over 2}dc\ \int_{2 vt - c \over 2}^{c}db\ {(vt + x)^2 \over 2 b c (2 vt - b - c)} +
\int_{vt-x \over 2}^{-2 x}dc\ \int_{2 vt - c \over 2}^{vt-c-x}db\ {(vt + x)^2 \over 2 b c (2 vt - b - c)}
\Bigg) =
$$

$$
{(vt+x)^2 \over 8 vt} {\rm{restr}}(-vt/2,x,-vt/3)
$$

$$
\Bigg( \pi^2 - 3 (\ln2)^2 - \ln 3 \ln {243 \over 63} + \big(\ln vt\big)^2 - \Li_2\bigg( {1 \over 9} \bigg) - 6 \Li_2\bigg( {2 \over 3} \bigg) + \big(\ln(vt - x)\big)^2 +
$$

$$
2 \Bigg[ \bigg( \arg \tanh {vt \over 2 vt+x} - \arg \tanh {x \over vt} \bigg) \ln 4 + \ln {vt+x \over vt-x} \ln(3 vt + x) - \ln(vt + x) \ln vt
$$

\begin{equation}\label{}
- \Li_2\bigg( {vt - x \over 4 vt} \bigg)
+ \Li_2\bigg( {vt - x \over 2 vt} \bigg)
+\Li_2\bigg( -{vt + x \over 2 vt} \bigg)
+\Li_2\bigg( {vt + x \over vt - x} \bigg)
- \Re\bigg( \Li_2\bigg( {1 \over 2} + {vt \over vt + x} \bigg) \bigg) \Bigg] \Bigg).
\end{equation}

$$
{\rm{restr}}(-vt/2,x,-vt/3)
$$

$$
\Bigg(
\int_{vt - x \over 2}^{-2 x}dc\ \int_{vt - c - x}^{c}db\ {b + c + 2 x \over 2 b c} +
\int_{-2 x}^{vt}dc\ \int_{x+vt}^{c}db\ {b + c + 2 x \over 2 b c} +
$$

$$
\int_{vt}^{vt-x}dc\ \int_{vt + x}^{2 vt-c}db\ {b + c + 2 x \over 2 b c}
\Bigg) =
$$

$$
{\rm{restr}}(-vt/2,x,-vt/3)
$$

$$
\Bigg(
{1 \over 2} \Big(-vt - x - x (\ln 2)^2 + (vt - x) (\ln (vt^2 - x^2) +
x \ln(vt - x)^2 - x (\ln vt)^2 \Big) +
$$

$$
x \ln(- 2 x) \bigg(1 + \ln {vt+x \over vt-x} \bigg)
- x \ln 2 \ln vt - vt \ln (vt+x) - x \ln (vt-x) \ln {vt + x \over 2 vt} +
$$

\begin{equation}\label{}
x \Li_2 \Big( { -2 x \over vt - x} \Big) - x \Li_2 \Big( {vt - x \over 2 vt} \Big)
\Bigg).
\end{equation}

$$
{\rm{restr}}(-vt/2,x,-vt/3) \int_{-2 x}^{vt-x}dc\ \int_{2 vt - c \over 2}^{vt+x}db\ {2 b^2 + 2 b (c - 2 vt) + (c-vt+x)^2 \over 2bc (b+c-2 vt)} =
$$

$$
{ {\rm{restr}}(-vt/2,x,-vt/3) \over vt }
$$

\begin{equation}\label{}
\Bigg( vt (vt + x) - {\pi^2 \over 48} (vt + x)^2 + vt x \ln {vt - x \over - 2 x} -
{1 \over 4} (vt - x)^2 \Li_2 {vt + x \over vt - x} \Bigg).
\end{equation}

$$
{\rm{restr}}(-vt/2,x,-vt/3) \int_{vt-x}^{2 vt}dc\ \int_{2 vt - c \over 2}^{2 vt-c}db\ {1\over c} =
$$

\begin{equation}\label{}
{\rm{restr}}(-vt/2,x,-vt/3) \Bigg(-{vt + x \over 2} + vt \ln {2 vt \over vt - x} \Bigg).
\end{equation}
\newpage

\newpage

The four integrals in the case $|x| \in [0,vt/3]$, when $x < 0$, are

\begin{figure}[h]
\centering
\includegraphics[width=0.99 \textwidth]{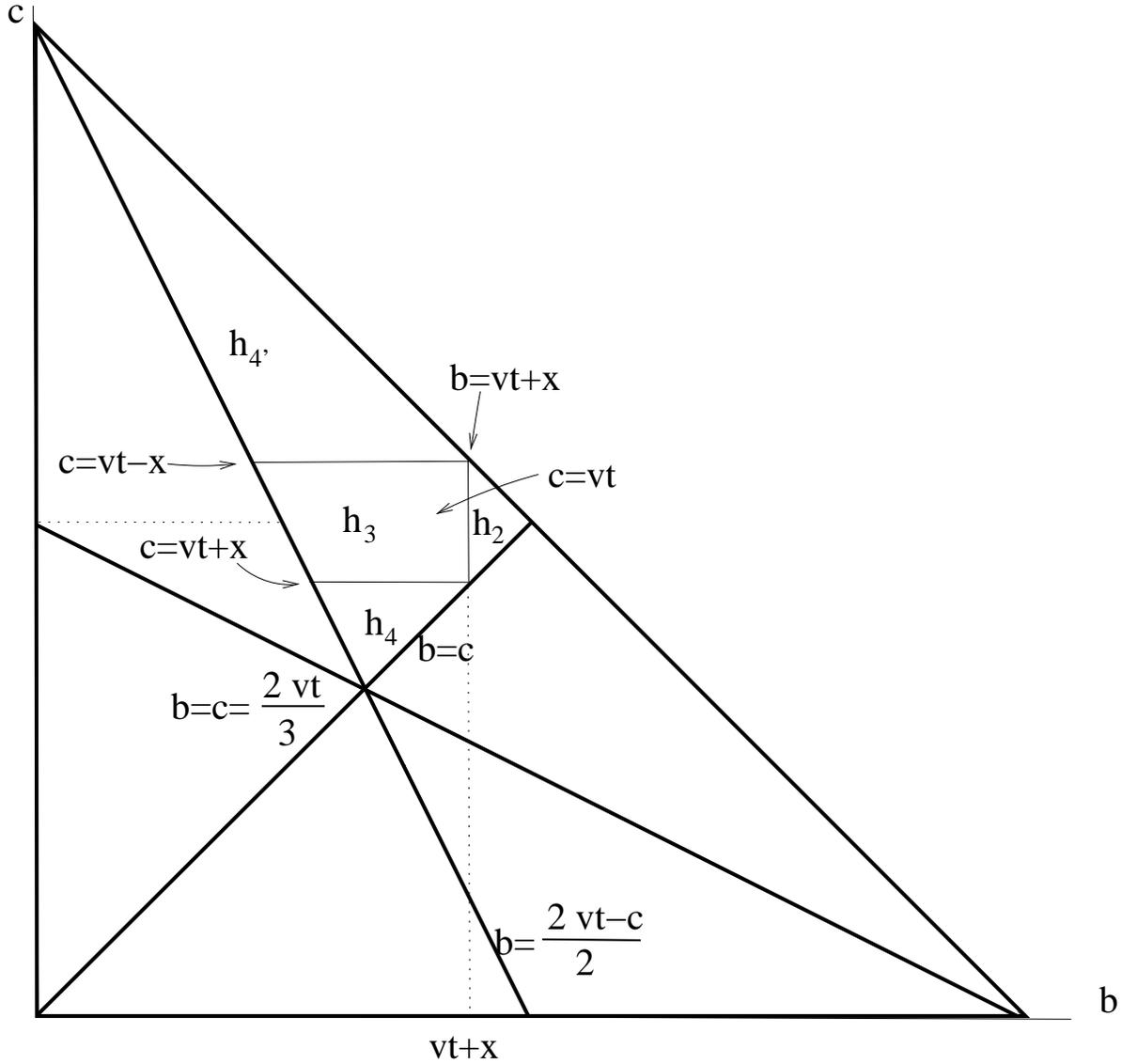}
\caption{\label{fig:tri-x-pequenos}The above tiling showing the different integrands and regions of integration in the $(b,c)$ plane is valid for $-vt/3 \leq x \leq 0$.}
\end{figure}

Due to the inequality $vt/3 \leq b+c-vt \leq vt$, the case $|x| \in [b+c-vt, vt]$, which calls for the integration of $h_1$, doesn't happen when $|x| \in [0, vt/3]$. Next, we have to ask if $x$ is smaller or greater than $-vt+b$. When it is smaller the integrand is ${b + c + 2 x \over 2 b c}$. When it is larger we have to distinguish the $c \in [2 vt/3, vt]$ and the $c \in [vt, 2 vt]$ cases. In the first case, when $x \in [-vt+b,-vt+c]$ the integrand is ${2 b^2 + 2 b (c - 2 vt) + (c-vt+x)^2 \over 2bc (b+c-2 vt)}$, and when $x \in [-vt+c,0]$ the integrand is ${2 b^2 + 2 b (c - 2 vt) + (c-vt)^2+x^2 \over bc (b+c-2 vt)}$. In the second case, when $x \in [-vt+b,vt-c]$ the integrand is ${2 b^2 + 2 b (c - 2 vt) + (c-vt+x)^2 \over 2bc (b+c-2 vt)}$, and when $x \in [vt-c,0]$ the integrand is $1/c$.

$$
{\rm{restr}}(-vt/3,x,0)
$$

$$
\Bigg(
\int_{vt + x}^{vt}dc\ \int_{vt + x}^{c}db\ {b + c + 2 x \over 2 b c} +
\int_{vt}^{vt-x}dc\ \int_{vt+x}^{2 vt-c}db\ {b + c + 2 x \over 2 b c}
\Bigg) =
$$

$$
{\rm{restr}}(-vt/3,x,0)
$$

$$
\Bigg(-x (\ln vt)^2 + {1 \over 2} x \ln {vt \over vt + x} \Big(-1 + \ln {vt \over vt + x} \Big) +
 \Big( {x \over 2} + x \ln {vt^2 - x^2 \over 2} \Big) \ln vt +
$$

\begin{equation}\label{}
{1 \over 12} \Big( \big( 12 + \pi^2 - 6 (\ln 2)^2 \big) x - 6 vt \ln (vt + x) +
    6 \big( vt + (-1 + \ln 4) x - 2 x \ln (vt + x) \big) \ln (vt - x) \Big) -
 x \Li_2 {vt - x \over 2 vt}
\Bigg).
\end{equation}

$$
{\rm{restr}}(-vt/3,x,0) \int_{vt+x}^{vt-x}dc\ \int_{2 vt - c \over 2}^{vt+x}db\ {2 b^2 + 2 b (c - 2 vt) + (c-vt+x)^2 \over 2bc (b+c-2 vt)} =
$$

$$
{{\rm{restr}}(-vt/3,x,0) \over 12 vt} \Bigg( -2 vt (\pi^2 vt + 12 x) +
3 \Big( (5 vt^2 + 2 vt x + x^2) (\ln (vt - x))^2 - 8 vt x \ln (vt + x) +
$$

$$
(vt - x)^2 (\ln (vt + x))^2 -
2 \ln (vt - x) \big( -2 vt x + (3 vt^2 + x^2) \ln (vt + x) \big) +
4 vt \ln (-2 x) (x + (vt + x) \ln {vt + x \over vt - x} \Big) +
$$

\begin{equation}\label{}
12 vt (vt + x) \Li_2 \Big( -{2 x \over vt - x} \Big) +
3 (5 vt^2 + 2 vt x + x^2) \Li_2 {vt + x \over vt - x} -
3 (vt + x)^2 \Re \bigg( \Li_2 \bigg(-1 + {2 vt \over vt + x} \bigg) \bigg) \Bigg).
\end{equation}

$$
{\rm{restr}}(-vt/3,x,0)
\int_{2 vt/3}^{vt+x}dc\ \int_{2 vt - c \over 2}^{c}db\ {b^2 + b (c - 2 vt) + (c-vt)^2 + x^2 \over bc (b+c-2 vt)} =
$$

$$
- {{\rm{restr}}(-vt/3,x,0) \over 12 vt}
\Bigg(
-6 vt^2 + (vt^2 + x^2) (\pi^2 - 3 (\ln 3)^2) - 18 vt x +
 6 (vt^2 + x^2) \ln {4 \over 3} \ln 3
$$

$$
 - 6 (vt^2 + x^2) \ln 2 \ln 2 vt +
 3 (vt^2 + x^2) (\ln 2 vt)^2 +
 6 \ln (-2 x) \Big( 2 vt x + (vt^2 + x^2) \ln {vt - x \over vt (vt + x)} \Big)
$$

$$
+ 3 \bigg( (vt^2 + x^2) (\ln (vt + x))^2 + 2 \ln (vt + x) \Big( -2 vt x + (vt^2 + x^2) \ln {-2 x \over vt-x} \Big)
$$

\begin{equation}\label{}
- 2 (vt^2 + x^2) \Big( \Li_2 \Big( -{1 \over 3} \Big) + 2 \Li_2 {2 \over 3} \Big)  \bigg) +
 6 (vt^2 + x^2) \bigg( \Li_2 \Big( {x \over vt} \Big) - \Li_2 \Big( {vt+x \over 2 vt} \Big) +
    \Li_2 \Big( {vt+x \over vt} \Big) \bigg)
\Bigg).
\end{equation}

\begin{equation}\label{roproj2.12}
{\rm{restr}}(-vt/3,x,0) \int_{vt-x}^{2 vt}dc\ \int_{2 vt - c \over 2}^{2 vt-c}db\ {1 \over c} =
{\rm{restr}}(-vt/3,x,0) \bigg(-{vt + x \over 2} - vt \ln {vt - x \over 2 vt} \bigg).
\end{equation}

The addition of the last 12 results is $\rho_{proj,2}(x,t)$. From this pdf

\begin{equation}\label{rho2}
\rho_{I,2}(x,t) = -{1 \over 2 \pi x} {d \over dx} \rho_{proj,2}(x,t),
\end{equation}
according to relation \eqref{inverseAbel.1}. The pdf $\rho_{I,2}$ is denoted as ``ro[2]'' in the file 4roc.nb in site \cite{RGPNgooglesite} and it is graphed in Fig. \ref{Comparacionchoques.1}.

\section{Appendix C. Computation}

There is a long way from expressions \eqref{decomposition.0}-\eqref{rho1.0} to the computation of actual values which allow to obtain the graphs of section \ref{Graphs}. The main difficulty is how to decide where to truncate the summations, and subsections C1-C3 of this appendix are devoted to that.

In order to compute $\rho_r$ and $\rho_I$, the summations in the collision expansion \eqref{PoissonExpansion} and in series \eqref{rorsc} and \eqref{ro1sc.1} over the indices $c$, $h$ and $m$ must be truncated. We need to estimate which the upper limits of the summations over the harmonics ($h$ index), the moments ($m$ index) and the collisions ($c$ index) should be. The estimation of the upper limit of $c$ is the only one which is easy. In subsections C1-C3 we estimate bounds for these summations such that $\rho_r$ and $\rho_I$ can be computed with a relative error below $10^{-3}$.

Expansion \eqref{PoissonExpansion} shows that the larger the time $t$, the more $\rho_c$'s are necessary to compute $\rho$ with a given accuracy. This, in turn, requires the computation of coefficients $C(c,m)$ for large $c$. Given that the computational power is limited, for large times the formulae found in this article have to yield to the Gaussian approximation. This is the topic of the last subsection, C6, of this appendix. The Gaussian approximation is used for times greater than $100/\lambda$. It is shown in subsection C6 that the relative error of the Gaussian approximation is larger, of about 0.01.

In sections C4-C5 comparison with the exact values 0 obtained in subsection \ref{Exact solution for two and more collisions} suggest that, with the truncations used in this appendix, the absolute error of the Fourier series approximation to $\rho_{rs}$ is bounded by $0.0027 \lambda/ v$, and that the absolute error of the Fourier series approximation to $\rho_{Is}$ is bounded by $0.0002 \lambda/ v$.

To compute the Fourier series coefficients we have used a high-level language and a rational approximation for pi, $\pi \approx {314159265358979323846 \over 100000000000000000000}$. In this way all calculations are done with rational numbers keeping all digits. Only when the calculation is done we ask the computer to convert the result to a decimal number.

A more sophisticated, precise and extensive discussion of the issues presented in this numerical section is certainly possible. But this article is not numerical.

\subsection{C1. Truncation of $c$}\label{sec:Truncation of c}

The infinite sum $\sum_{c=1}^\infty$ which appears in the Poisson expansion \eqref{PoissonExpansion} has to be substituted by $\sum_{c=1}^{c_{max}}$ when performing a calculation. We choose a $c_{max}$ such that at most a prescribed fraction $\epsilon_r$ of the norm of $\rho$ would be neglected if $\sum_{c=1}^\infty$ were substituted without further ado by $\sum_{c=1}^{c_{max}}$. That is, $c_{max}$ is defined by the inequalities

\begin{equation}\label{}
{1 \over e^{\lambda t} - 1} \sum_{c=c_{max}+1}^\infty {(\lambda t)^c \over c!} \leq \epsilon_r \leq {1 \over e^{\lambda t} - 1} \sum_{c=c_{max}}^\infty {(\lambda t)^c \over c!}\ \ \ \ \rightarrow\ \ \ c_{max}(\lambda t, \varepsilon_r).
\end{equation}
The denominators are $e^{\lambda t} - 1$ because in the Poisson expansion \eqref{PoissonExpansion} the sum starts at $c=1$. Once $c_{max}(\lambda t, \varepsilon_r)$ is obtained the Poisson expansion \eqref{PoissonExpansion} is substituted by

\begin{equation}\label{Truncationc}
\rho(\vec r, t) = {1 \over \sum_{j=1}^{c_{max}(\lambda t, \varepsilon_r)} \l( (\lambda t)^j/j! \r)} \sum_{c=1}^{c_{max}(\lambda t, \varepsilon_r)} {(\lambda t)^c \over c!} \rho_{c}(\vec r, t)
\end{equation}
so that the truncated expression remains normalized.

The following data have been obtained numerically: $c_{max}(0.1, 10^{-3}) = 2,\ c_{max}(1, 10^{-3}) = 5,\ c_{max}(10, 10^{-3}) = 21,$ $c_{max}(50, 10^{-3}) = 73,\ c_{max}(100, 10^{-3}) = 132$. That
the norm of a truncation in the number of collisions departs from 1 by less than $10^{-3}$ does not guarantee that at any given point $x$ the truncated pdf and the true one differ by less than 1
part in 1,000, but it is reasonable to expect that it cannot differ by a lot more. Hence, from now on we suppose then, that with these $c$'s the relative error of approximation \eqref{Truncationc}
is going to be bounded by $10^{-3}$.

\subsection{C2. Truncation of $h$}

For each $c$ we have computed enough Fourier coefficients $FS(\rho_{Is,c})(h)$ so that the ratio of the smallest (which is $FS(\rho_{Is,c})(hmax)$ or $FS(\rho_{Is,c})(hmax-1)$) to the largest (which is $FS(\rho_{Is,c})(1)$) is at most $10^{-3}$. We expect, then, the relative error of the series to be bounded by $10^{-3}$ in the region around the origin where the amplitudes, except for the ones which correspond to large frequencies, mostly add up. In the tail, however, the argument in the preceding sentence no longer holds, and the relative error can be larger than $10^{-3}$. A very similar discussion would hold for $\rho_{rs,c}$.

\subsection{C3. Truncation of $m$}\label{sec:Truncation of m}

Similarly to \eqref{FSro1} we define the Fourier series coefficient $FS(\rho_{rs,c})(h)$ by

$$
FS(\rho_{rs,c})(h) \equiv \sum_{m=0}^\infty (-\pi^2)^{m} {(2m+1) c! \over (2m+c)!} C(c,m) (h^2)^m =
$$

$$
\left\{ (-\pi^2)^{m} {(2m+1) c! \over (2m+c)!} C(c,m) \right\}_{m=0}^\infty \cdot \left\{ (h^2)^m \right\}_{m=0}^\infty  \approx
$$

\begin{equation}\label{SFc}
\left\{ (-\pi^2)^{m} {(2m+1) c! \over (2m+c)!} C(c,m) \right\}_{m=0}^{mMax} \cdot \left\{ (h^2)^m \right\}_{m=0}^{mMax}\ \ \ {\rm{for}}\ \ \ h=1,2,3,...,
\end{equation}
which we have written in a manner that suggests a scalar product. The advantage of this is that, for each $c$, $\left\{ (-\pi^2)^{m} {(2m+1) c! \over (2m+c)!} C(c,m) \right\}_{m=0}^{mMax}$ may be computed first for a large $mMax$. This sequence may be stored and then multiplied scalarly by the sequence $\left\{ (h^2)^m \right\}_{m=0}^{mmax}$, where $mmax$ might be smaller than $mMax$ if not much accuracy is needed.

The trouble with the terms $\left\{ (-\pi^2)^{m} {(2m+1) c! \over (2m+c)!} C(c,m) (h^2)^m \right\}_{m=0}^\infty$ is that $m$ has to be large for the terms to start becoming small. For a given $hmax$, $mmax$ will be found by numerical experimentation. That is, for each $hmax$, $mmax$ is such that the numerical behaviour of the Fourier coefficients $FS(\rho_{rs,c})(h)$ is good for $h < hmax$. This numerical behaviour is quite characteristic and shown in Fig. \ref{SFrs}. For a fixed $mmax$, these coefficients begin to diverge wildly and rather suddenly beyond a certain $h$. Conservatively, we choose $mmax$ so that $hmax$ starts to diverge a few units after $hmax$.

In this paragraph an idea is given to guide the numerical search for $mMax$. Suppose, to start, that $c=1$. Since $C(1,m)<1$ (see its definition \eqref{C(c,m)def} or appendix A), in order to find when the terms become small it is enough to think about
$\sum_{m=m_{Max}+1}^\infty {\pi^{2m} h^{2m}/((2m)!)} \overset{\rm{Stirling}}{\approx} \sum_{m=m_{max}+1}^\infty {(e \pi)^{2m} h^{2m}/((2m)^{2m})}$. The rhs is a geometric series whose ratio becomes
$\leq 1/2$ when $m \geq 9 h$. When $c>1$ we still use the previous estimate, because $(\pi^2)^{m} (2m+1) c! C(c,m) (h^2)^m/((2m+c)!)$ decays faster than $\pi^{2m} h^{2m}/(2m)!$. The result
$m \geq 9 h$ gives us an idea of where to start in order to find the $mMax$ necessary for a given $hmax$. Indeed, we have found in the examples discussed in subsections C4 and C5 that a $mMax$ of
5-10 times $hmax$ is necessary. A very similar discussion would hold for $\rho_{Is,c}$.

\subsection{C4. Computation of $\rho_{rs,c}$}\label{Computation of rorc}

In this section we use expansion \eqref{rorsc} to find $\rho_{rs,c}$ for $c=0,1,2,3,10$ and make some comments by which the reader will become familiar with the truncations of \eqref{rorsc} as well as with other aspects of the functions $\rho_{rs,c}$. The radial pdf conditional to 0 collisions is not in the collision expansion \eqref{PoissonExpansion}, but is given here for completeness. Since $C(0,m) = {1 \over 2m +1}$,
$$
\rho_{rs,0}(x,t) = {1 \over 2 v t}\ \sum_{h=-\infty}^{+\infty} \sum_{m=0}^\infty {(-1)^m (\pi h)^{2m} \over (2m)!} e^{i {\pi h \over v t} x} =
{1 \over 2 v t}\ \sum_{h=-\infty}^{+\infty} \cos \pi h\ e^{i {\pi h \over v t} x} =
$$

\begin{equation}\label{}
{1 \over 2 v t}\ \sum_{h=-\infty}^{+\infty} (-1)^h \cos{\pi h \over v t} x =
{1 \over 2 v t}\ \sum_{h=-\infty}^{+\infty} e^{-i {2 \pi h \over 2 v t} v t} \cos{\pi h \over v t} x,
\end{equation}
which the reader will recognize as the Fourier series expansion of a comb function whose spikes lie a distance $2 v t$ apart and which has been shifted a distance $vt$ from the origin. Its restriction to the $[-vt,+vt]$ interval is one half Dirac delta function at each of its end points.

The radial pdf conditional to 1 collision is

\begin{equation}\label{_{rs}1}
\rho_{rs,1}(x,t) = {1 \over 2 v t}\ \sum_{h=-\infty}^{+\infty} \sum_{m=0}^\infty {(-1)^m (\pi h)^{2m} \over (2m)!} \bigg( \sum_{i=0}^m {1 \over (2 i+1) (2 (m-i)+1)} \bigg) \cos {\pi h \over v t} x,
\end{equation}
which is the Fourier series of (see eq. \eqref{rho1})

\begin{equation}\label{rho1.1}
\rho_{rs,1}(x,t) = {1 \over 2} 4 \pi x^2 \rho_{I,1}(x,t) = {x \over 2 (vt)^2} \ln {v t+x \over v t-x}
\end{equation}
if its values on its support $x \in [-vt, +vt]$ are repeated all over the real line so as to make it a periodic function.

We have set $mMax=500$ and found that the Fourier series coefficients start to diverge when $h=115$. We have summed expression \eqref{_{rs}1} up to the 50-th and the 100-th harmonic. The former sum is compared with the exact expression \eqref{rho1.1} in Fig. \ref{Comparacion1choque}. The approximation provided by the latter sum is too good to be clearly distinguishable from the exact graph provided by expression \eqref{rho1.1}.

\begin{figure}[h]
\centering
  \includegraphics[width=0.5 \textwidth]{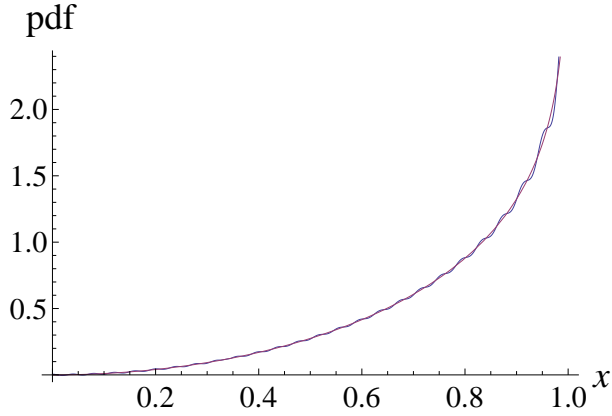}
  \caption{Comparison of the Fourier series $\rho_{rs,1}(x,1/v)$ with $\lambda=1$, summed up to the 50-th harmonic and the 500-th moment, with the exact expression for $v t = 1$. When the Fourier series is summed up to the 100-th harmonic the graph becomes barely distinguishable from the exact expression.}\label{Comparacion1choque}
\end{figure}

The radial pdf conditional to 2 collisions is

$$
\rho_{rs,2}(x,t) = {1 \over 2 v t}\ \sum_{h=-\infty}^{+\infty} 2 \sum_{m=0}^\infty (-\pi^2)^{m} {(2m+1) \over (2m+2)!} C(2,m) (h^2)^m e^{i {\pi h \over v t} x} =
$$

\begin{equation}\label{_{rs}2}
{1 \over v t}\ \sum_{h=-\infty}^{+\infty} \left[ \sum_{m=0}^\infty  {(-\pi^2)^{m} \over (2m+2) (2m)!} C(2,m) (h^2)^m \right] \cos {\pi h \over v t} x.
\end{equation}
This expression is compared in Fig. \ref{Comparacion2choques} with the $\rho_{rs,2}(x,t)$ obtained from $\rho_{proj,2}(x,t)$ (see Appendix B after expression \eqref{roproj2.12}) by formula \eqref{inverseAbel.1}, $\rho_{rs,2}(x,t) = -x {d \over dx} \rho_{proj,2}(x,t)$.

\begin{figure}[h]
\centering
  \includegraphics[width=0.5 \textwidth]{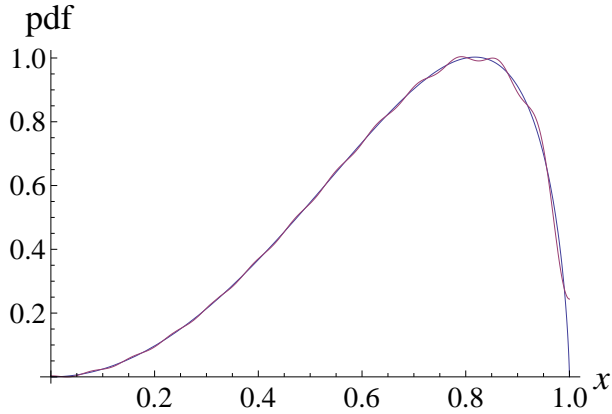}
  \caption{Comparison of the Fourier series $\rho_{rs,2}(x,1/v)$ with $\lambda=1$, summed up to the 25-th harmonic ($mMax=500$), with the exact $\rho_{rs,2}(x,t)$ expression for $v t = 1$. When the Fourier series is summed up to the 50-th harmonic the graph becomes barely distinguishable from the exact expression.}\label{Comparacion2choques}
\end{figure}

\bigskip
The examination of the general form of the Fourier series coefficients \eqref{SFc} of $\rho_{rs,c}$ suggests that the larger the $c$, the faster they fall off as $h$ increases, and thus, as implied by the caption of Fig. \ref{Comparacion2choques}, 50 terms or less should give a good fit for all $c>2$. Indeed, this is confirmed when the Fourier series coefficients for $\rho_{rs,1},\ \rho_{rs,2},\ \rho_{rs,3}$ and $\rho_{rs,10}$ are compared. See Fig. \ref{SFrs} for the first three. As of $\rho_{rs,10}$, its first 12 coefficients are $\{ 0.329598, -0.468924, -0.340769, -0.028119, 0.010381, -0.00274265, 0.000735024,\\ -0.000204368, 0.000056844, -0.0000143126, 2.1067 \times 10^{-6}, 1.0685 \times 10^{-6} \}$, they decay a lot faster than the ones of $\rho_{rs,1},\ \rho_{rs,2}$ and $\rho_{rs,3}$.

\begin{figure}[h]
\centering
  \includegraphics[width=0.45 \textwidth]{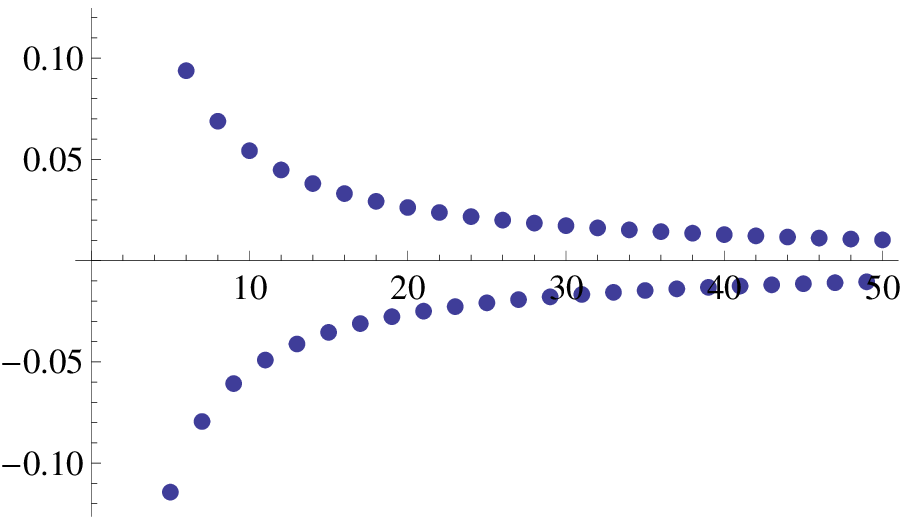}\hspace{01cm}\includegraphics[width=0.45 \textwidth]{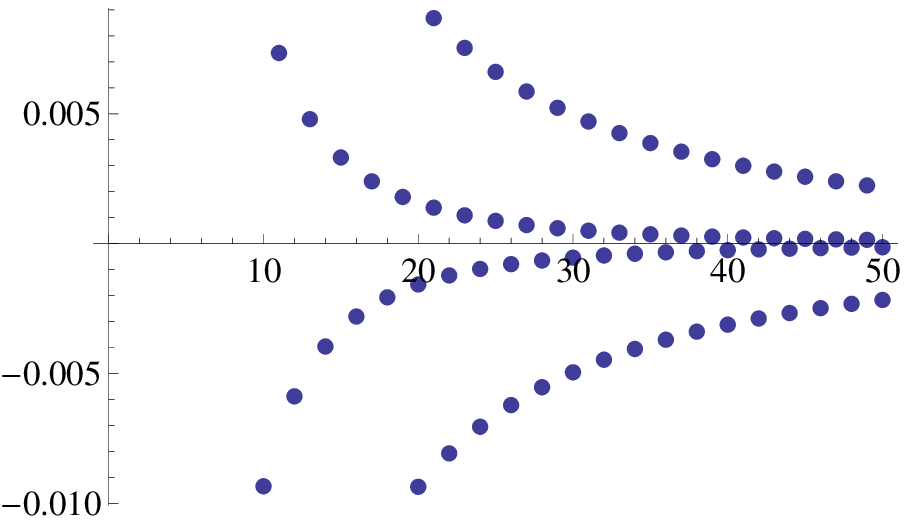}
  \caption{To the left, the first 50 Fourier series coefficients for $\rho_{rs,1}$ when $vt=1/2$. To the right, the first 50 Fourier series coefficients for $\rho_{rs,2}$ (the outer points) and $\rho_{rs,3}$ (the inner points) when $vt=1/2$. Notice the different scales in each graph.}\label{SFrs}
\end{figure}

When $c>2$ we don't have exact $\rho_{rs,c}$ to compare with, but, as we have just argued, we can obtain good approximations for $\rho_{rs,c}$ with fewer Fourier series coefficients. A test to see how good the approximations (see Fig. \ref{rors3y10}) to $\rho_{rs,3}$ (with 100 Fourier series coefficients) and $\rho_{rs,10}$ (with 25 Fourier series coefficients) are is to evaluate them at $x = 0, vt$. They should be 0, as shown in subsection \ref{Exact solution for two and more collisions}, but table 2 shows that they are not.

In Fig. \ref{Comparacion2choques}, which depicts the Fourier series approximation to $\rho_{rs,2}(x,1/v)$ with 25 harmonics, the largest departure between the truncated Fourier series and the exact solution takes place at $x=1$. If we suppose that this happens also for $c = 3$ and $c = 10$, then the values 0.013 and $-1.4 \times 10^{-7}$ given in table 2 not only give an idea of the error, but are actually an upper bound of the absolute error of the truncated Fourier series $\rho_{rs,3}$ and $\rho_{rs,10}$, respectively, when $t=1/v$.

The information of the previous paragraph leads us to make two assumptions: 1) The largest error of the truncated Fourier series for $\rho_{rs,c}$ always takes place at $x = \pm vt$. 2) Using the truncations of subsections \ref{sec:Truncation of c}-\ref{sec:Truncation of m}, the error at $x = \pm vt$ diminishes with increasing $c$. Then 0.013 is an upper bound for the errors of $\rho_{rs,c}(x,1/v)$ for $c \geq 3$ and, from expansion \eqref{PoissonExpansion} and scaling relation \eqref{ScalingRhoc2}, the absolute error of $\rho_{rs}$, to be denoted by err$_{rs}$, satisfies

\begin{equation}\label{err_rs}
{\rm{err}}_{rs} \leq {1 \over e^{\lambda t} - 1} \sum_{c=3}^\infty {(\lambda t)^c \over c!} {0.013 \over vt} = {e^{\lambda t} - 1 - \lambda t - {(\lambda t)^2 \over 2} \over \lambda t (e^{\lambda t} - 1)}\ {\lambda\ 0.013 \over v}.
\end{equation}
This bound is 0 when $t=0$ and tends to 0 when $t \to \infty$. The first factor of the last term of inequality \eqref{err_rs} has a maximum of about 0.202351 at $\lambda t \approx 3.02377$. Thus we conclude that

\begin{equation}\label{}
{\rm{err}}_{rs} \leq 0.202351\ {\lambda\ 0.013 \over v}  \leq \approx 0.0027 {\lambda \over v}
\end{equation}
for all times.

\begin{figure}[h]
\centering
  \includegraphics[width=0.45 \textwidth]{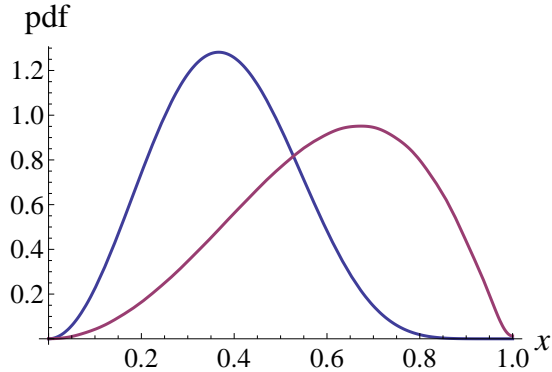}\hspace{01cm}
  \caption{Graphs of $\rho_{rs,3}(x,1/v)$ and $\rho_{rs,10}(x,1/v)$ (the taller curve) with 100 and 25 Fourier series coefficients, and $mMax=500$ and $mMax=125$, respectively.}\label{rors3y10}
\end{figure}

\subsection{C5. Computation of $\rho_{I,c}$}\label{Computation of roIc}

So far everything has seemed to be going rather well with the pdf's $\rho_{rs,c}$. The match between the exact solutions and the truncated Fourier series was good and the required number of harmonics diminished with $c$. However, the pdf that we are after is usually not $\rho_{r}$ but $\rho$. To obtain $\rho$ we may, following eqs. \eqref{rorsc}, \eqref{PoissonExpansion}, \eqref{rhors} and \eqref{rhoI}, divide the $\rho_{rs,c}$ functions by $2 \pi r^2$. But this is numerically not satisfactory, as advertised after eq. \eqref{rI}. Indeed, one can see in Fig.  \ref{Comparacionchoques.1} that oscillations appear in the vicinity of 0 which will not go away by increasing the frequency cut off.

\begin{figure}[h]
\centering
  \includegraphics[width=0.45 \textwidth]{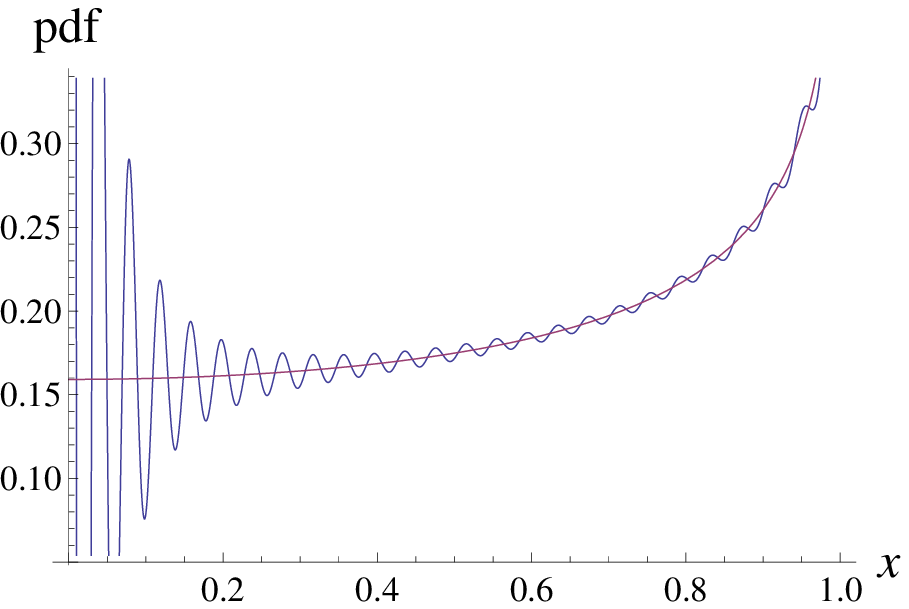}\hspace{01cm}\includegraphics[width=0.45 \textwidth]{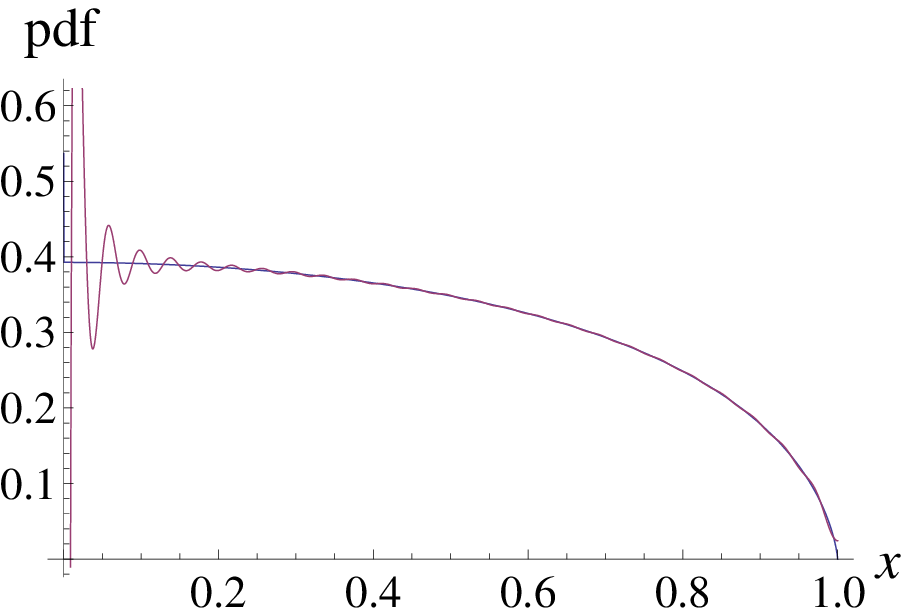}
  \caption{To the left, the exact $\rho_{Is,1}(x,1/v)$ (one half of expression \eqref{rho1}) and the Fourier series \eqref{rorsc} for $\rho_{rs,1}(x,1/v)$ with 50 harmonics divided by $4 \pi x^2$. To the right, the exact $\rho_{Is,2}(x,1/v) = {1 \over 2} \rho_{I,2}(x,1/v)$ (where $\rho_{I,2}$ is derived in Appendix B and is denoted by ``ro[2]'' in the file 4roc.nb in \cite{RGPNgooglesite}) and the Fourier series \eqref{rorsc} for $\rho_{rs,2}(x,1/v)$ with 50 harmonics divided by $4 \pi x^2$.}\label{Comparacionchoques.1}
\end{figure}

\begin{figure}[h]
\centering
  \includegraphics[width=0.45 \textwidth]{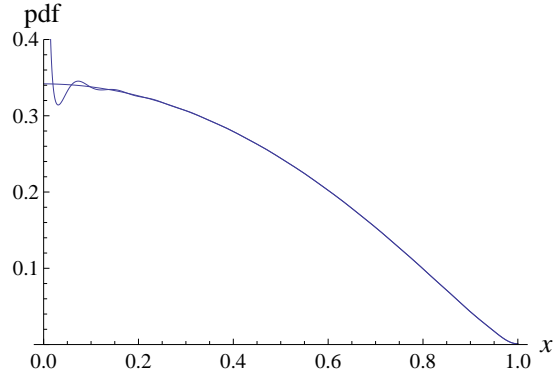}
  \caption{The graph of $\rho_{rs,3}(x,1/v)/(4 \pi x^2)$ with 25 Fourier series coefficients and $mMax=125$ misbehaves between $x=0$ and about $x = 0.1 v t$. But in the graph of $\rho_{Is,3}(x,1/v)$ with 25 Fourier series coefficients the oscillations have disappeared.}\label{rors3YroIs3bis}
\end{figure}

Note that the problem of oscillations is less serious for $c=2$ than for $c=1$. Indeed, already for $c=3$ Fig. \ref{rors3YroIs3bis} shows that the problem is mild enough to be solved by various numerical methods. But a better method is the one given in the second half of section \ref{Fourier series of r and I}. It follows from expansion \eqref{Desarrollo} that

\begin{equation}\label{ro1sc.2}
\rho_{Is,3}(x,t) = {1 \over (vt)^3} \sum_{h=0}^{\infty} FS(\rho_{Is,3})(h) \cos {\pi h \over v t} x.
\end{equation}
We need to find $\langle r^{-2} \rangle_{\rho_{3}}(1/v)$ to be able to compute $FS(\rho_{Is,3})(h)$ for $h=0,1,2,3,...$ (see definitions \eqref{FSro1.0} and \eqref{FSro1}). It follows from equation \eqref{limFS} and from definition \eqref{FS_0} that

\begin{equation}\label{menosdos}
\langle r^{-2} \rangle_{\rho_{3}}(1/v) = - \lim_{h \to \infty}
\sum_{m=1}^\infty { (-\pi^2)^m 3!\ C(3,m-1) \over 2 m (2m-2+3)!} \cdot h^{2m}
\approx 5.4408
\end{equation}
as seen in Fig. \ref{ArmonicosSFIs3}. We can now use formula \eqref{ro1sc.1} to compute $\rho_{Is,3}(x,1/v)$. Its graph is depicted in Fig. \ref{rors3YroIs3bis}.

\begin{figure}[h]
\centering
  \includegraphics[width=0.45 \textwidth]{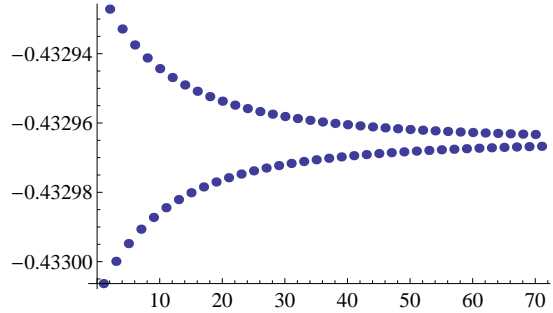}
  \caption{Plot of the coefficients $FS_0(\rho_{Is,3})(h)$ from $h=30$ to $h=100$. They tend to -0.432965..., therefore $\langle r^{-2} \rangle_{\rho_{3}}(1/v) \approx -4 \pi (-0.432965) = 5.4408$}\label{ArmonicosSFIs3}
\end{figure}
\noindent
%
%

The examination of the general form of the Fourier series coefficients \eqref{ro1sc.1} of $\rho_{Is,c}$ suggests that the larger the $c$, the faster they fall off as $h$ increases, and thus, as implied by the left figure of Fig. \ref{Comparacion2choquesI}, 100 terms or less should give a good fit for all $c>2$. Indeed, this is confirmed when the Fourier series coefficients for $\rho_{Is,1},\ \rho_{Is,2},\ \rho_{Is,3}$ and $\rho_{Is,10}$ are compared. See Fig. \ref{fig:SF1sc} for the first three. As of $\rho_{Is,10}$, its first 12 coefficients are $\{ 0.9236, 0.3286, 0.0331, -0.0045, 0.00068, -0.000114, 0.000018,-1.26 \times 10^{-6}, -1.36 \times 10^{-6}, 1.30\times 10^{-6}, -9.10 \times 10^{-7}, 5.77 \times 10^{-7} \}$, they become negligible a lot faster than the ones of $\rho_{Is,1},\ \rho_{Is,2}$ and $\rho_{Is,3}$.

\begin{figure}[h]
\centering
  \includegraphics[width=0.45 \textwidth]{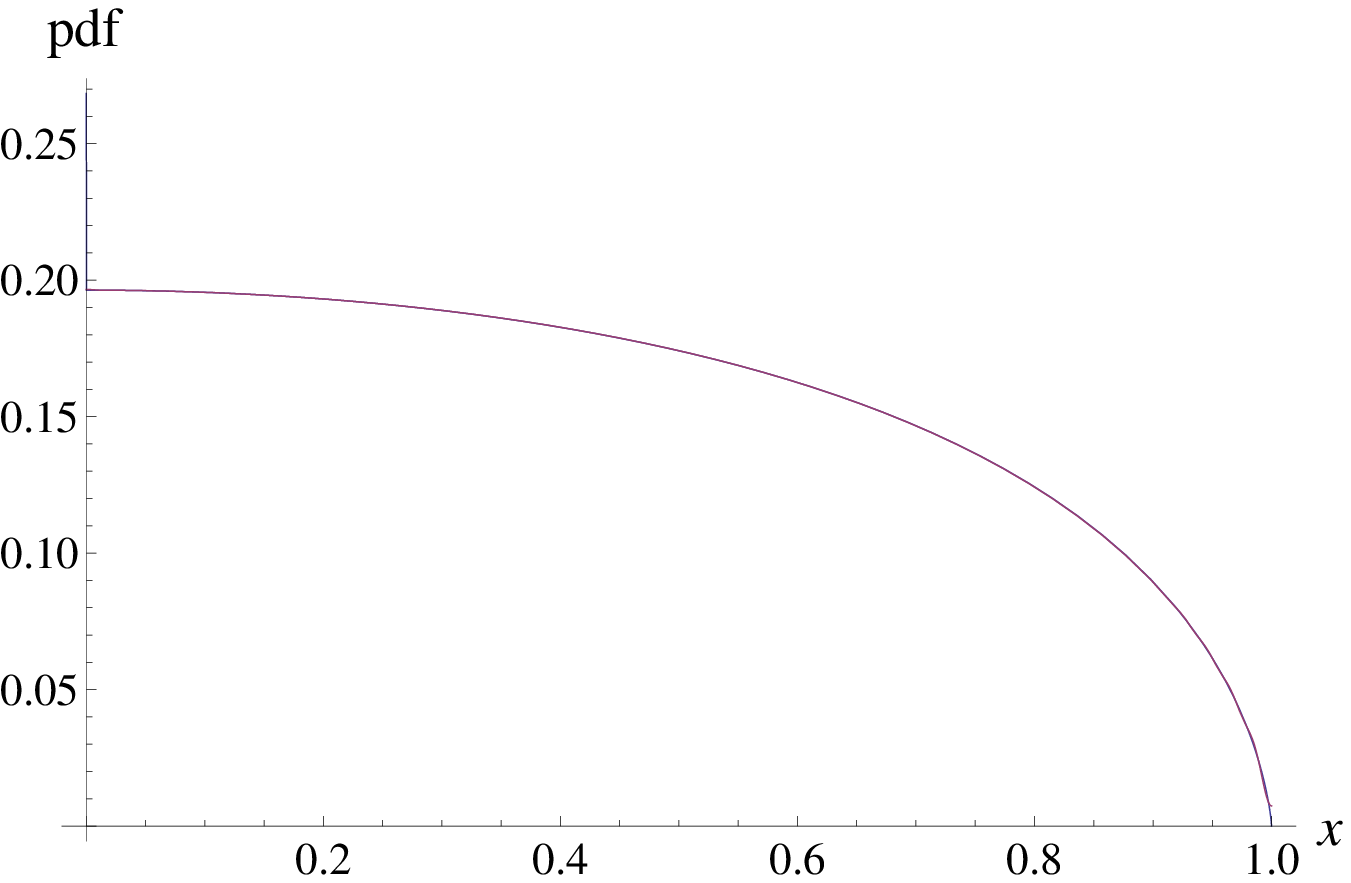} \hspace{1cm} \includegraphics[width=0.45 \textwidth]{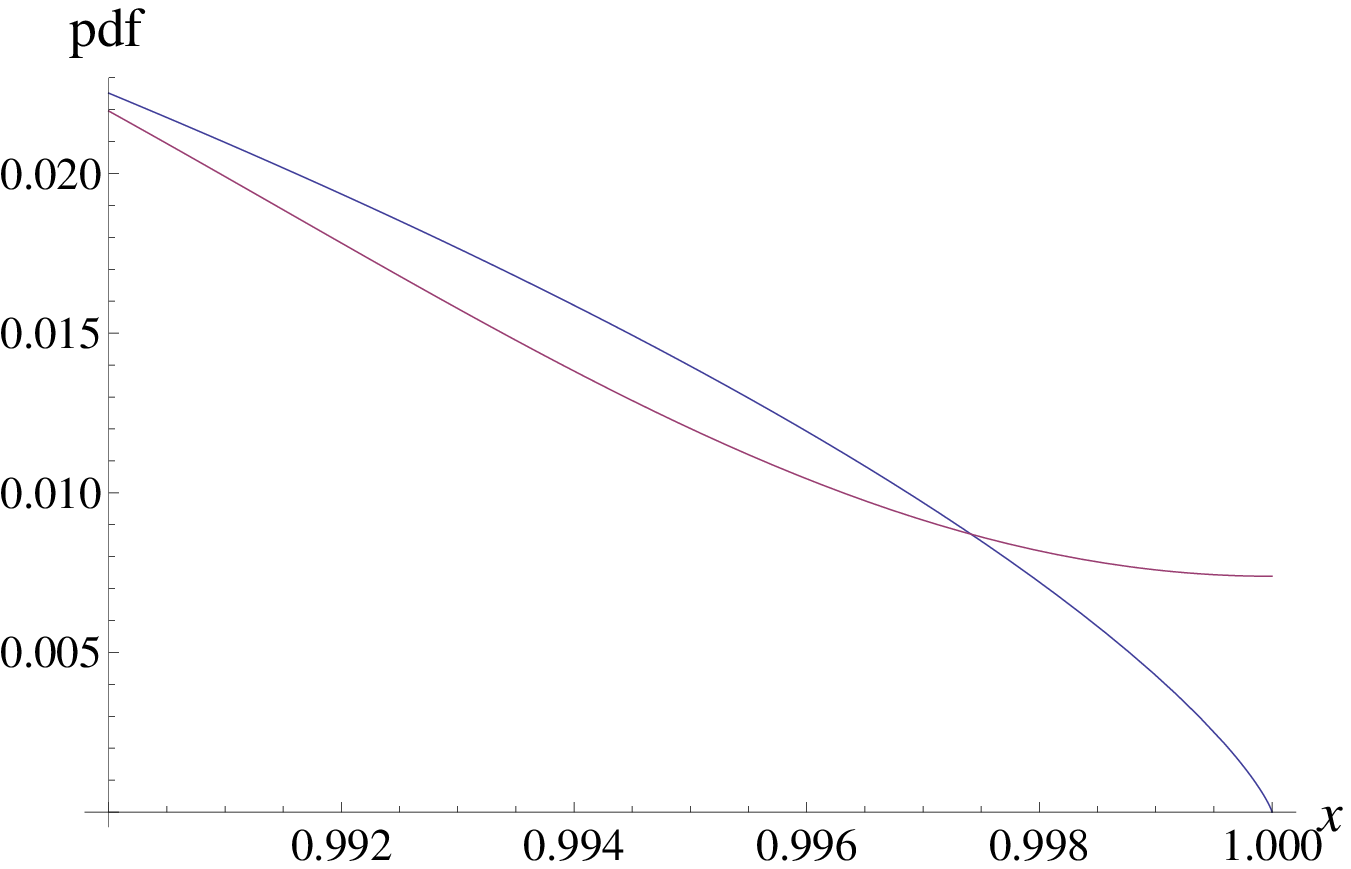}
  \caption{Comparison of the Fourier series $\rho_{Is,2}(x,1/v)$ with $\lambda=1$, summed up to the 100-th harmonic ($mMax=500$), with the exact $\rho_{Is,2}(x,t)$ expression for $v t = 1$ (half of expression \eqref{rho2}, because of definition \eqref{rhoIs}). On the left figure both graphs are indistinguishable from each other, but on the right figure one can see that the largest difference between both functions happens at $x=1$, where the Fourier series takes the value 0.007385...}\label{Comparacion2choquesI}
\end{figure}

\begin{figure}[h]
\centering
  \includegraphics[width=0.33 \textwidth]{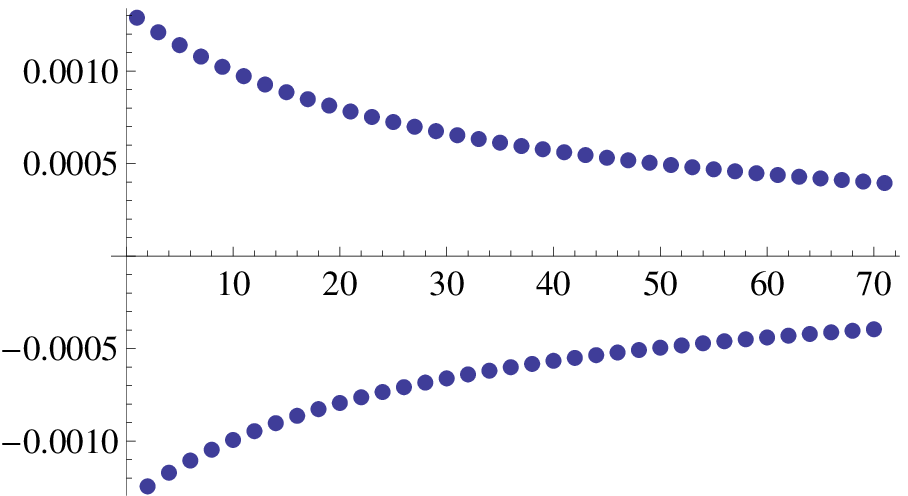}\includegraphics[width=0.33 \textwidth]{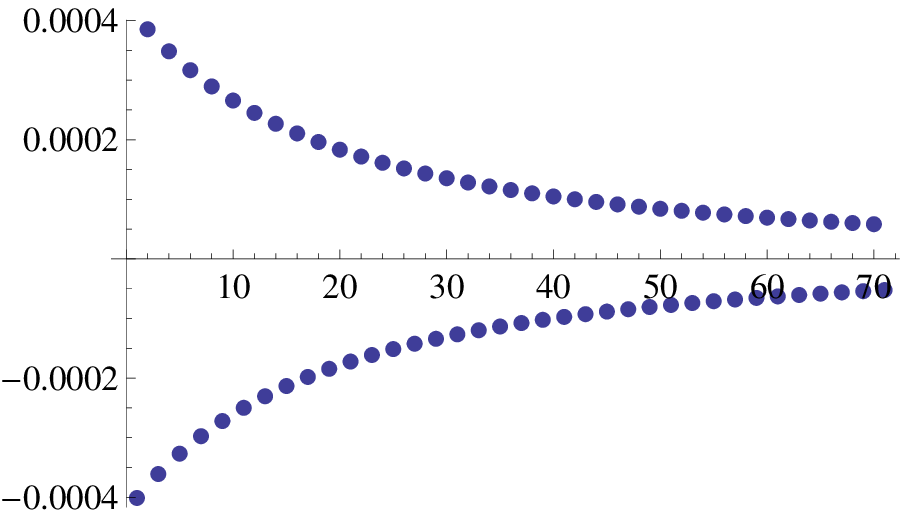}\includegraphics[width=0.33 \textwidth]{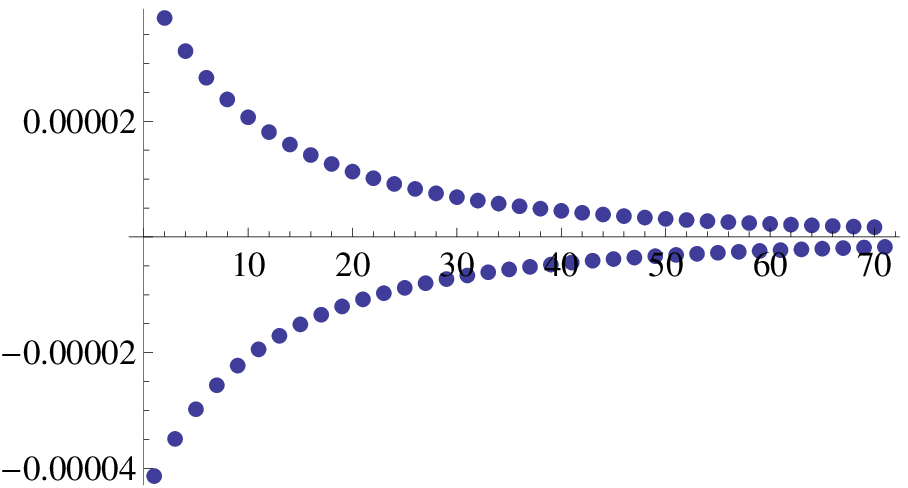}
  \caption{From left to right, the 30-th to 100-th Fourier series coefficients for $\rho_{Is,1}$, $\rho_{Is,2}$ and $\rho_{Is,3}$, when $vt=1$. Notice the different scales in each graph.}\label{fig:SF1sc}
\end{figure}

\begin{tabular} { | p {1in} | p {1in} | p {1in} | p {1in} | p {1in} | }
\hline
\multicolumn{5} { | c | }{Table 2: Errors of the Fourier series for $t=1/v$ truncated to $h$ harmonics}\\
\hline
\hline
& \multicolumn{2} { | c | }{$x=0$} & \multicolumn{2} { | c | }{$x=\pm vt$} \\
\hline
& $c=3$ & $c=10$ & $c=3$ & $c=10$ \\
\hline
$\rho_{rs,c}(x,1/v)$ & ($h = 100$) & $(h = 25)$ & ($h = 100$) & $(h = 25)$ \\
& 0.00041 & $-1.75 \times 10^{-8}$ & 0.013 & $-1.4 \times 10^{-7}$ \\
\hline
$\rho_{Is,c}(x,1/v)$ &  &  & $(h = 25)$ & $(h = 12)$  \\
&  &  & 0.001 & $-10^{-6}$ \\
\hline
$\rho_{Is,c}(x,1/v)$ &  &  & $(h = 100)$ & $(h = 25)$ \\
&  &  & 0.0001 & $-10^{-8}$ \\
\hline
\end{tabular}
\bigskip


As in subsection \ref{Computation of rorc} we use the results of subsection \ref{Exact solution for two and more collisions} to estimate the error when the Fourier series \eqref{ro1sc.1} is truncated. In particular it follows from Proposition 5.2 \eqref{Proposition 2} that $\rho_{Is,3}(\pm vt,1/v)=0$ and $\rho_{Is,10}(\pm vt,1/v)=0$, but table 2 shows that with the value for $\langle r^{-2} \rangle_{\rho_{Is,3}}$ given by the limit \eqref{menosdos}, the value given by the truncated Fourier series is not 0. We remind the reader that $\rho_{Is,3}(x,1/v)$ is plotted in Fig. \ref{rors3YroIs3bis}.

In Fig. \ref{Comparacion2choquesI}, which depicts the Fourier series approximation to $\rho_{Is,2}(x,1/v)$ with 100 harmonics, the largest departure between the truncated Fourier series and the exact solution takes place at $x=1$. If we suppose that this happens also for $c = 3$ and $c = 10$, then the figures shown in table 2 not only give an idea of the error, but are actually an upper bound of the absolute error of the truncated Fourier series $\rho_{Is,3}$ (with 25 and 100 harmonics) and $\rho_{Is,10}$ (with 12 and 25 harmonics) when $t=1/v$.

The information of the previous paragraph leads us to make two assumptions: 1) The largest error of the truncated Fourier series for $\rho_{Is,c}$ always takes place at $x = \pm vt$. 2) With the truncations of subsections \ref{sec:Truncation of c}-\ref{sec:Truncation of m}, the error at $x = \pm vt$ diminishes with increasing $c$. Then 0.001 is an upper bound for the errors at $x=1$ of $\rho_{Is,c}(x,1/v)$ for $c \geq 3$ and, from expansion \eqref{PoissonExpansion} and scaling relation \eqref{ScalingRhoc2} we can do the same analysis as at the end of subsection \ref{Computation of rorc} to conclude that the absolute error of $\rho_{Is}$, to be denoted by err$_{Is}$, satisfies

\begin{equation}\label{err_Is}
{\rm{err}}_{Is} \leq {e^{\lambda t} - 1 - \lambda t - {(\lambda t)^2 \over 2} \over \lambda t (e^{\lambda t} - 1)}\ {\lambda\ 0.001 \over v}.
\end{equation}
This bound is 0 when $t=0$ and tends to 0 when $t \to \infty$. Again as at the end of subsection \ref{Computation of rorc} we conclude that

\begin{equation}\label{}
{\rm{err}}_{Is} \leq \approx 0.0002 {\lambda \over v}
\end{equation}
for all times.

\begin{figure}[h]
\centering
  \includegraphics[width=0.45 \textwidth]{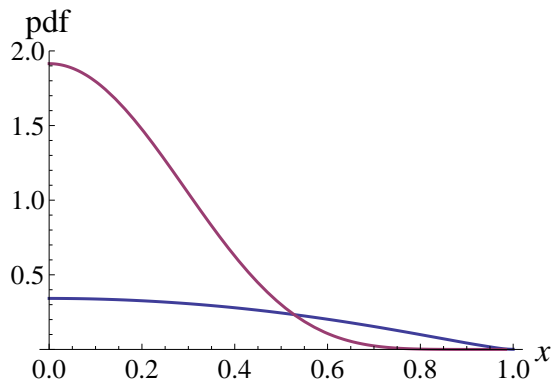}
  \caption{Graphs of $\rho_{Is,3}(x,1/v)$ and $\rho_{Is,10}(x,1/v)$ (big, bell shaped). They are the graphs of Fig. \ref{rors3y10} divided by $4 \pi x^2$. While the curves in Fig. \ref{rors3y10} had to cover the same area, here this needs not to be so.}\label{roIs3YroIs10}
\end{figure}

Bear in mind that, according to definitions \eqref{rhors} and \eqref{rhoIs}, the graphs of $\rho_{rs,c}$ and $\rho_{Is,c}$ depicted in this section are half as high as the graphs of $\rho_{r,c}$ and $\rho_{I,c}$ over the same domain.

\subsection{C6. Asymptotics}\label{Asymptotics}

It has been shown \cite{Shlesinger1974,Stadje1987,Boguna1998,Kolesnik2008,RGPN2012cap} that as $t \to \infty$, $\rho$ tends to a Gaussian. Two parameters are needed to specify a Gaussian of zero mean: its norm and its variance, $\langle r^0 \rangle$ and $\langle r^2 \rangle$, respectively. In three dimensions a Gaussian with given norm and variance is:

\begin{equation}\label{}
{\langle r^0 \rangle \over \Big( {2 \over 3} \pi {\langle r^2 \rangle \over \langle r^0 \rangle} \Big)^{3/2}}
\exp -{3 \over 2} {r^2 \over \langle r^2 \rangle/\langle r^0 \rangle}.
\end{equation}
In our case the norm of the limiting Gaussian is (see decomposition \eqref{decomposition.0})

\begin{equation}\label{}
\langle r^0 \rangle(t) = 1-e^{-\lambda t}.
\end{equation}
From formula (29) of reference \cite{RGPN1995} we know that the second moment of the limiting Gaussian is

\begin{equation}\label{}
\langle r^2 \rangle(t) = 2 \Big( {v \over \lambda} \Big)^2 \big( e^{-\lambda t} - 1 + \lambda t \big) - e^{-\lambda t} (v t)^2,
\end{equation}
where we have subtracted the contribution due to the expanding spherical shell from the said formula. Therefore

\begin{equation}\label{GaussianAprox}
\rho(r,t) \underset{t \to \infty}{\approx}
{1-e^{-\lambda t} \over \Big( {2 \over 3} \pi {\langle r^2 \rangle(t) \over 1-e^{-\lambda t}} \Big)^{3/2}}
\exp -{3 \over 2} {  r^2 \over { \langle r^2 \rangle(t) \over 1-e^{-\lambda t} }  }.
\end{equation}

\begin{figure}[h]
\centering
\includegraphics[width=0.5 \textwidth]{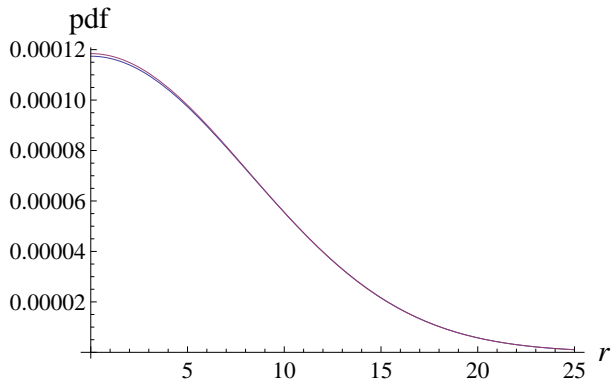}\label{fig:Asint1}
\caption{Plots of the Gaussian approximation and the expansion with 132 collisions for $\lambda t = 100$. The Gaussian approximation is the taller curve.}
\end{figure}

At $(r,\lambda t) = (0,100)$ the ratio of the Gaussian approximation to the Fourier series with 132 harmonics is 1.0088. Since we have taken 132 harmonics, as stated at the beginning of this Appendix and as developed in subsections C1-C3, the relative error of the series \eqref{ro1sc.1} at $(r,\lambda t) = (0,100)$ is at most $10^{-3}$ in the region around $r=0$, which is almost an order of magnitude less than 0.0088. We may then take the expansion \eqref{ro1sc.1} with 132 collisions as exact relative to the Gaussian approximation in the said region and conclude that if, for $\lambda t>100$, we use the Gaussian approximation, then the relative error in the central region will always be less than 1$\%$, because $1.0088 < 1.01$.

This is a setback from the $10^{-3}$ relative error for $\lambda t>100$, but pushing the transition to the Gaussian approximation to a larger time where the Gaussian approximation became significantly better would require the computation of a couple of hundred more collections of Fourier coefficients, up to 300 collisions or more. Such a numerical effort is beyond the scope of this paper.

\bigskip

\newpage



\end{document}